\documentclass[aps,prl,reprint,superscriptaddress]{revtex4-2}

\usepackage[utf8]{inputenc}
\usepackage[T1]{fontenc}
\usepackage{lmodern}
\usepackage{amsmath,amssymb,bm}
\usepackage{graphicx}
\usepackage{siunitx}

\graphicspath{{fig/}}

\begin{document}
	
	\title{Tailoring complex coupling for efficient synchronization of semiconductor lasers}
	
	\author{Nathan Vigne}
	\affiliation{Department of Applied Physics, Yale University, New Haven, Connecticut 06520, USA
	}%
	
	\author{Amit Pando}
	\affiliation{Department of Physics of Complex Systems, Weizmann Institute of Science, Rehovot 7610001, Israel
	}%
	
	\author{Li-Li Ye}
	\affiliation{School of Electrical, Computer and Energy Engineering, Arizona State University, Tempe, Arizona 85287, USA
	}%
	
	\author{Mehmet B\"ut\"un}
	\affiliation{Department of Applied Physics, Yale University, New Haven, Connecticut 06520, USA
	}%
	
	\author{KyeoReh Lee}
	\affiliation{Department of Applied Physics, Yale University, New Haven, Connecticut 06520, USA
	}%
	
	\author{Ying-Cheng Lai}
	\affiliation{School of Electrical, Computer and Energy Engineering, Arizona State University, Tempe, Arizona 85287, USA
	}%
	\affiliation{Department of Physics, Arizona State University, Tempe, Arizona 85287, USA
	}%
	
	\author{Nir Davidson}
	\affiliation{Department of Physics of Complex Systems, Weizmann Institute of Science, Rehovot 7610001, Israel
	}%
	
	\author{Fan-Yi Lin}
	\affiliation{Institute of Photonics Technologies, Department of Electrical Engineering, National Tsing Hua University, Hsinchu 30013, Taiwan
	}%
	
	\author{Hui Cao}
	\affiliation{Department of Applied Physics, Yale University, New Haven, Connecticut 06520, USA
	}%
	\email[Corresponding Author]{hui.cao@yale.edu}

	\date{\today}
	
	\begin{abstract}
		A grand challenge for laser synchronization is the intrinsic frequency disorder. We build an experimental platform for programming the complex coupling coefficients of semiconductor lasers to promote frequency and phase locking. The coupling amplitude between every pair of lasers in an array is tuned to scale with their frequency difference.  Coupling phases are scanned to meet the phase matching conditions for constructive interference of coupled fields. The minimum coupling budget for global synchronization is significantly reduced, as shown experimentally for three vertical-cavity surface-emitting lasers with intrinsic frequency detuning. The customized coupling offers an efficient approach to stable synchronization of semiconductor laser arrays for coherent beam combining. 
	\end{abstract}
	
	
	\maketitle
	
	
	Synchronization of coupled nonlinear oscillators has provided a platform for understanding the emergence of collective behavior in large ensembles of interacting dynamical systems \cite{arenas2008synchronization, dorogovtsev2008critical}. One important application is laser synchronization, i.e., frequency and phase locking of coupled lasers \cite{Soriano2013ComplexPhotonics, Ohtsubo2017, LangKobayashi1980, Mulet2002Bidirectional, Wunsche2005DelayCoupled}, to generate an intense beam for material processing and lighting \cite{Fan2005, LiuBraiman2013CBC, Kapon1984}. More recently, coupled-laser networks have been utilized for photonic reservoir computing and deep learning \cite{Skalli2022VCSELNeuromorphic, chen2023deep}. 
	A longstanding obstacle for global synchronization is intrinsic frequency disorder, i.e., free-running frequencies of uncoupled lasers are different due to small variations in geometry and material composition. Semiconductor lasers are compact and a large array can be fabricated in a single chip, but their intrinsic amplitude-phase coupling may destabilize phase-locked states \cite{Henry1982, WinfulWang1988}. Over the past few decades, diverse coupling schemes and network topologies have been explored for laser synchronization, from nearest-neighbor evanescent coupling \cite{Kapon1984, WinfulWang1988, orenstein1992Large2DVCSEL} to long-range diffractive coupling \cite{liu2004synchronization, Brunner:15, Nair2018WeaklyCoupled}, and global all-to-all coupling \cite{kozyreff2000global, zamora2010crowd, Nixon2013GeometricFrustration, Argyris:16}. However, most coupling architectures prescribe the network topology independently of the frequency detuning of individual lasers \cite{bourmpos2012sensitivity, xiang2016synchronization}. Here our aim is to tailor the laser coupling based on frequency disorder for efficient synchronization. 
	
	For a given frequency disorder, what would be an efficient way of allocating a finite coupling budget for global synchronization of nonlinear oscillators? For non-identical Kuramoto oscillators, theoretical and numerical studies have shown that optimized network topologies acquire positive correlations between coupling strength and frequency detuning  \cite{brede2008synchrony, kelly2011topology, skardal2014optimal, lei2023new, Pando2026Proportional}. However, these studies are not directly applicable to semiconductor lasers, as they neglect amplitude-phase coupling and time-delayed interactions, which have significant impact on synchronization dynamics.    
	
	In addition to coupling amplitude, coupling phase plays a key role in laser synchronization. However, existing studies on synchrony-optimized networks are focused mainly on real coupling coefficients. For delay-coupled semiconductor lasers, it has been shown that tuning the global coupling phase through small delay adjustment can strongly modify the synchronization dynamics, locking regions, and selected collective modes \cite{Wunsche2005DelayCoupled, Erzgraber2005Bifurcation, nixon2009PhaseLocking, nixon2011synchronized, nixon2012controlling}. More generally, network synchronization is sensitive to phase frustration \cite{SakaguchiKuramoto1986, Nixon2013GeometricFrustration}. Theoretical studies reveal that chaos synchronization of identical oscillators strongly depends on individual coupling phases  \cite{flunkert2012chaos}. Nevertheless, independent tuning of coupling phases between all laser pairs has not been realized, and how it will affect steady-state synchronization of semiconductor lasers with frequency disorder is not known. 	
	
	Here, we set up an experimental platform for simultaneous, independent control of reciprocal pairwise coupling amplitude and phase of semiconductor lasers in a chip. It allows us to program the coupling amplitude for each pair of lasers based on their intrinsic frequency difference. In particular, we realize the `proportional' coupling, i.e., the coupling amplitude for any two lasers is proportional to their free-running frequency difference. It significantly lowers the coupling budget needed for synchronization of three vertical cavity surface-emitting lasers (VCSELs), compared to uniform coupling amplitude for every pair. The synchronization behavior changes dramatically from sequential locking of lasers with increasing frequency detuning for uniform coupling to simultaneous locking of all lasers with proportional coupling. We further scan the phases of complex coupling coefficients for individual pairs and observe a strong variation of synchronization threshold. The coupling phases not only determine the relative phases of locked lasers but also set the locked frequency for all lasers. 
	Adjusting the coupling phases allows matching the phase of each laser with the phases of coupled fields from other lasers. These fields will interfere constructively, fully utilizing the limited coupling budget for laser synchronization. Our conclusion is general and applicable to different types of lasers and nonlinear oscillators. 
	
	\label{sec:coupling_platform}
	
	\begin{figure}[ht!]
		\centering
		\includegraphics[width=\linewidth]{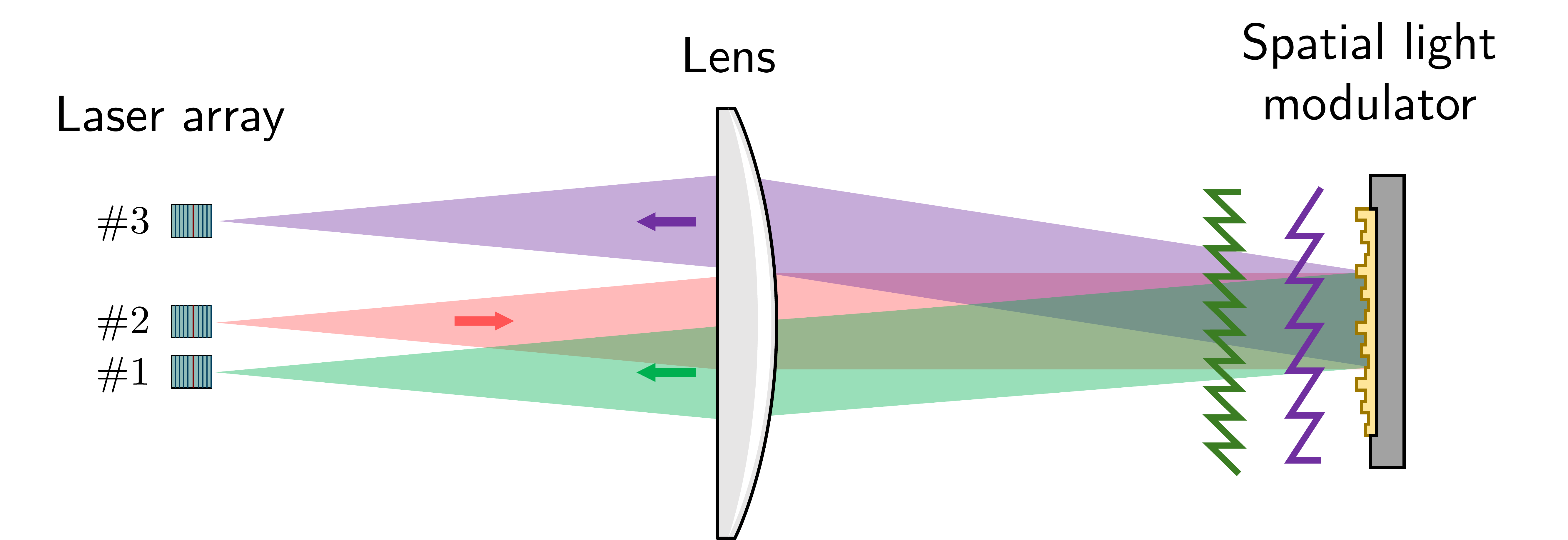}
		\caption{{\bf Programmable coupling of lasers in a chip.} Schematic of laser coupling by a spatial light modulator (SLM) at the Fourier plane of the laser chip. For illustration, two phase gratings with different periods diffract emission from laser 2 (red) to lasers 1 (green) and 3 (purple) simultaneously.}
		\label{fig:Schematic}
	\end{figure}
	
	We conduct a synchronization experiment on a one-dimensional (1D) array of VCSELs from the VI-system \textit{GmbH} \cite{VigneSM26}. The electric current injected into each laser can be separately controlled by a dedicated current driver. VCSEL supports a single longitudinal mode due to short cavity length. At the pump current of 1 mA, lasing occurs only in the fundamental transverse mode with linear polarization. The lasing frequency can be fine-tuned with slight adjustment of electric current. The spacing between neighboring VCSELs is 250 $\mathrm{\mu m}$, thus evanescent wave coupling is negligible. 
	
	To program the complex coupling coefficients of lasers in a chip, we use a spatial light modulator (SLM) as a reconfigurable diffractive element in an external cavity. As shown schematically in Fig.~\ref{fig:Schematic}, 
	multiple phase gratings of varying periods on the SLM simultaneously couple all laser pairs with prescribed amplitudes and phases. Using a first-order reflection, we can control the global coupling strength.
	The coupling is reciprocal: the complex coupling coefficient from laser $m$ to $n$ is equal to that from laser $n$ to $m$: $\kappa_{nm} = \kappa_{mn}$. For three VCSELs, there are three independent coupling coefficients: $\kappa_{12}$, $\kappa_{23}$, $\kappa_{13}$. To realize separate control of them, we select three lasers with different spatial spacing. To avoid self-coupling that may induce lasing instability, the SLM does not provide direct reflection of the emission from one laser back to itself. For precise control of coupling amplitude and phase, we optimize the SLM pattern using a gradient descent algorithm and an experimentally calibrated forward model \cite{VigneSM26}. The coupling budget is given by the mean coupling strength per laser, $b_{\mathrm{tot}}=(1/M)\sum_{n,m}{|\kappa_{nm}|}$, where $M$ is the total number of lasers \cite{Arenas2008SynchronizationComplexNetworks, MikaberidzeTaylor2025Emergent}. 
	
	To monitor laser synchronization, we track both frequency locking and phase locking \cite{VigneSM26}. To observe frequency locking, we perform heterodyne measurement of three VCSELs with a reference laser (same type of VCSEL). This allows us to track frequency differences of three coupled lasers and observe when they lase at the same frequency (frequency locking). To evaluate phase locking, we image the far-field emission from three VCSELs with a camera.  Once the lasers are frequency and phase locked, their emission will interfere coherently in the far field and produce stable fringes. The fringe contrast reflects the degree of phase locking. The synchronization threshold is defined as the lowest coupling budget $b_{\mathrm{tot}}$ at which the lasers are both frequency and phase locked. 
	
\begin{figure}[ht!]
		\centering
		\includegraphics[width=\linewidth]{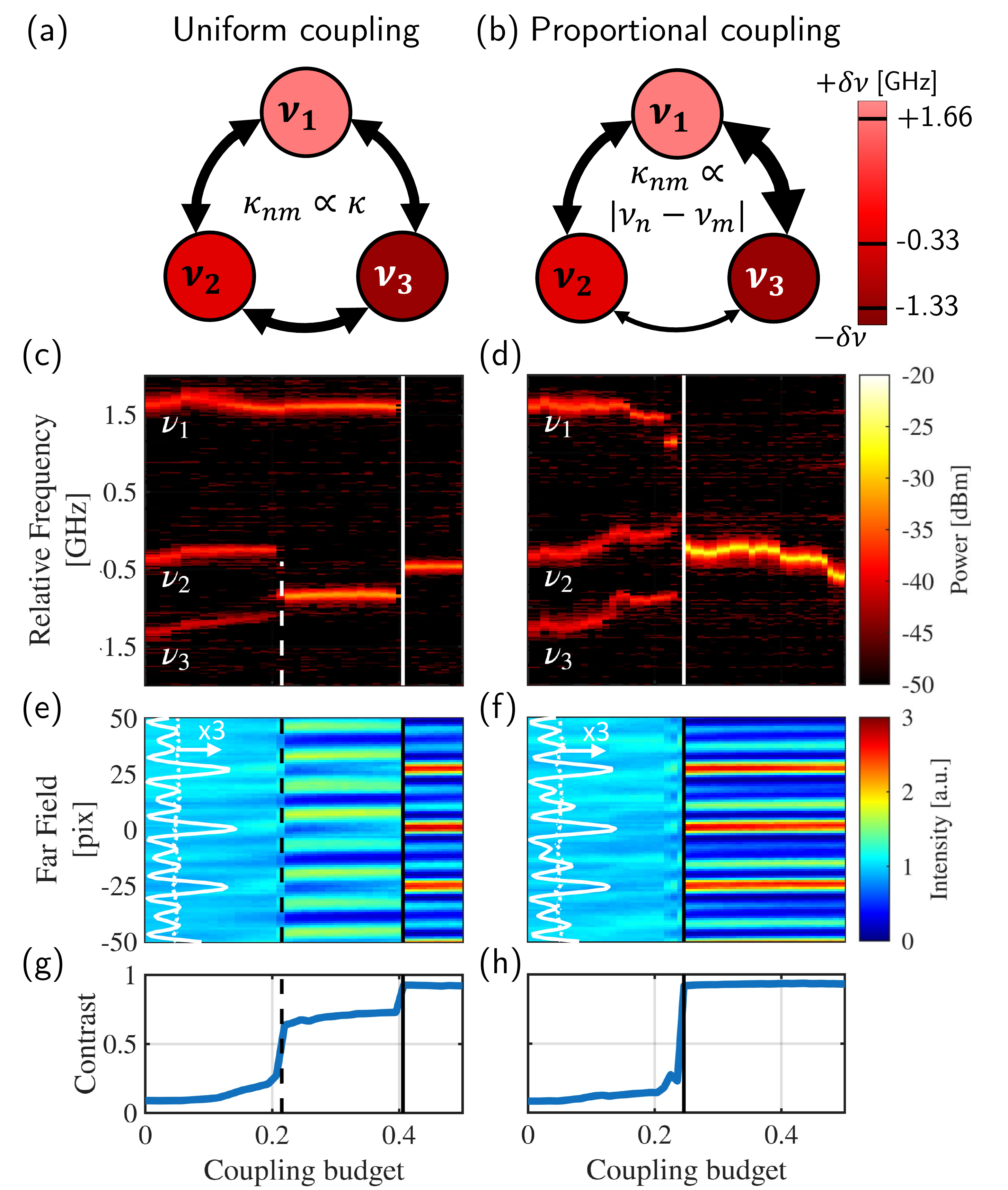}
		\caption{{\bf Frequency and phase locking with uniform and proportional coupling.} (a, b) Schematic representation of uniform (left) and proportional (right) coupling of three lasers. The free-running frequencies of uncoupled lasers are detuned by $\Delta \nu_1 = 1.66$ GHz, $\Delta \nu_2 = -0.33$ GHz, $\Delta \nu_3 = - 1.33$ GHz from the mean. The coupling phases are fixed. (c, d) Relative frequencies of three lasers with increasing coupling budget. The solid white line shows the frequency locking threshold. The dashed line indicates the two laser locking thresholds for uniform coupling. (e, f) Far-field intensity distribution with increasing coupling budget. Phase-locking threshold (marked by black vertical line) coincides with frequency-locking threshold. The inset white curves show the far-field intensity profile without coupling (dotted) and at the phase-locking threshold (solid). (g, h) Far-field fringe contrast showing two-step phase locking with uniform coupling and a single-step locking with proportional coupling.}
		\label{fig:amplitude_results}
	\end{figure}

	
	We compare the synchronization thresholds of proportional coupling and uniform coupling for three VCSELs. In the former, the pairwise coupling amplitude is proportional to the free-running frequency difference $|\kappa_{nm}|~\propto~|\nu_{n}~-~\nu_{m}|$. This is an extension of the two-oscillator linear relation between detuning and coupling strength needed for synchronization \cite{Mulet2002Bidirectional, adler1946Injection}. While in the latter, every laser pair has identical coupling amplitude $|\kappa_{nm}|~=~ \kappa$, see Fig.~\ref{fig:amplitude_results}(a, b). 
	
	The heterodyne spectra in Figs.~\ref{fig:amplitude_results}(c, d) reveal the onset of frequency locking in both coupling schemes. 
	With uniform coupling, the three lasers exhibit a sequential frequency locking in Fig.~\ref{fig:amplitude_results}(c). As the coupling budget increases, lasers 2 and 3 with the smallest frequency difference are locked first, then the most-detuned laser 1 is locked in frequency. 
	In contrast, for the proportional coupling in Fig.~\ref{fig:amplitude_results}(d), the frequencies of three lasers simultaneously collapsed to a common frequency at a critical coupling budget. The frequency locking threshold is lower than that for uniform coupling. This difference is attributed to a ``smart'' allocation of limited coupling budget, as laser pairs with larger frequency differences need stronger mutual injection to lock than pairs with smaller differences. 
	
	To acquire phase-locking signature, we examine the far-field emission patterns of three VCSELs in Figs.~\ref{fig:amplitude_results}(e,f). 
	In the case of uniform coupling, the interference fringes are first noticeable when lasers 2 and 3 are frequency locked (dashed vertical line), and become more pronounced when laser 1 is locked (solid vertical line). The fringe contrast ($C$) in Fig.~\ref{fig:amplitude_results}(g) exhibits two consecutive jumps, corresponding to phase locking of two lasers (2, 3) and then three lasers (1, 2, 3). In the case of proportional coupling, Fig.~\ref{fig:amplitude_results}(h), the fringes suddenly appear and display high contrast when three lasers are simultaneously frequency locked. Such high contrast ($C>0.9$) is reached only when three lasers have a time-invariant phase relation. We set the phase-locking threshold at $C=0.9$. It coincides with the frequency-locking threshold for both uniform and proportional coupling, revealing frequency locking and phase locking occur simultaneously. 
	As shown in the inset of Figs.~\ref{fig:amplitude_results}(e-f), the far-field peak intensity with locking (white solid line) is about 3 times of the mean intensity without locking (white dashed line), confirming the brightness enhancement by three locked lasers. 
	
	Proportional coupling not only lowers the coupling budget for laser synchronization from uniform coupling, but also changes the sequential synchronization to simultaneous synchronization. With identical coupling amplitude for all laser pairs, the uniform coupling scheme wastes some coupling budget on less-detuned pairs. By providing stronger coupling to more-detuned laser pairs, the proportional coupling scheme facilitates a collective transition of all lasers with different frequencies to the synchronized state.

\begin{figure}[ht!]
		\centering
		\includegraphics[width=\linewidth]{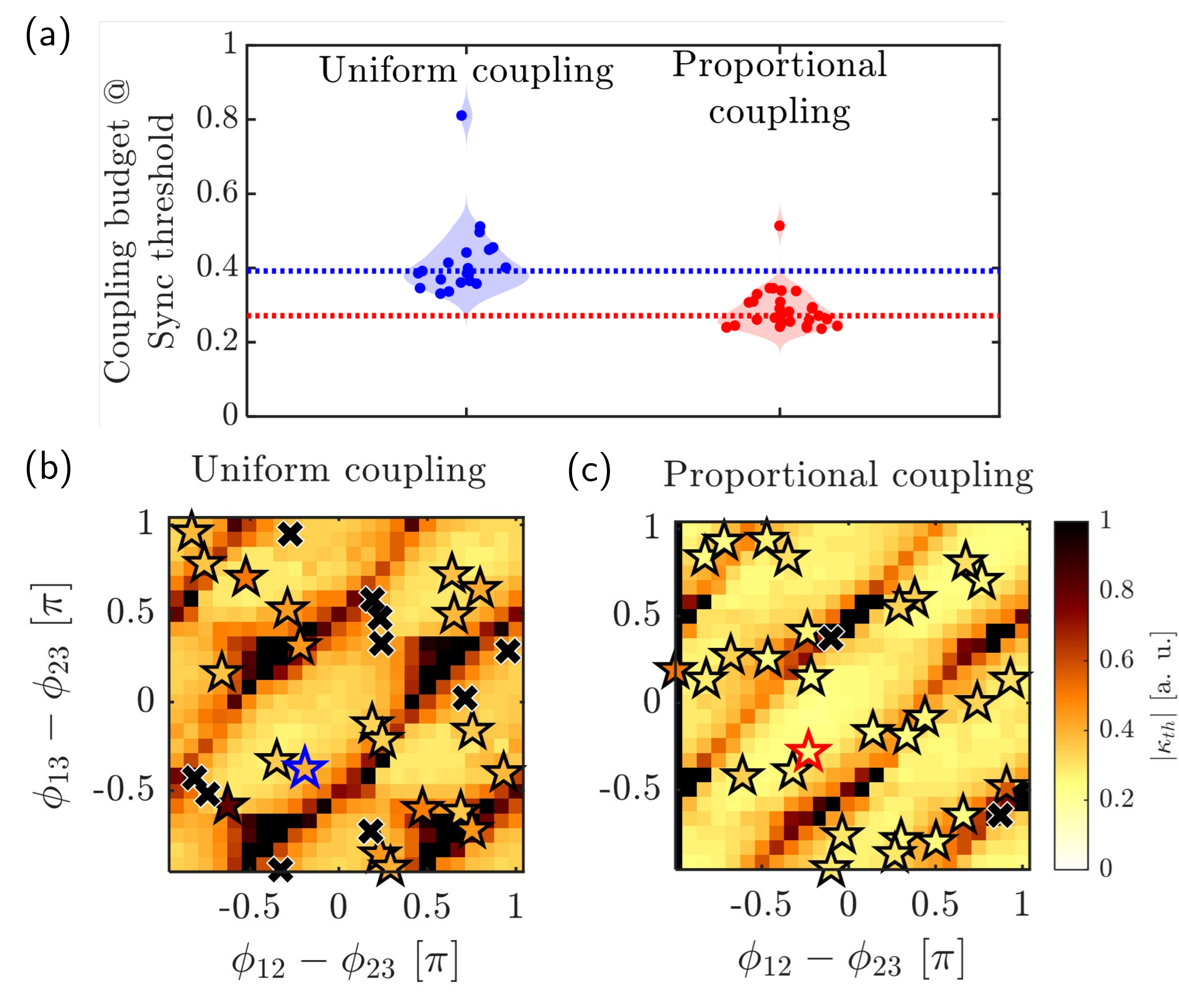}
		\caption{{\bf Coupling-phase dependence of synchronization threshold.} Experimentally measured synchronization thresholds for uniform and proportional coupling of three lasers (identical to those in Fig.~\ref{fig:amplitude_results}) with different coupling phases in a swarmchart (dots), with colored background representing the probability density. The median threshold value is marked by a dotted line. (b, c) Numerically calculated coupling budget at synchronization threshold as a function of relative phases $\phi_{12} - \phi_{23}$ and $\phi_{13} - \phi_{23}$ of complex coupling coefficients for uniform and proportional coupling, respectively. Experimental data are overlaid as stars with inner color representing the measured coupling budget at synchronization threshold. Highlighted blue and red stars correspond to Fig.~\ref{fig:amplitude_results} data. Black crosses correspond to no synchronization up to the maximum experimental coupling budget. The numerical coupling budget is normalized to $K_0 = \qty{62.18}{\giga\hertz}$, which represents the highest possible coupling rate. The relative coupling phases are measured experimentally \cite{VigneSM26}.}
		\label{fig:TH_Sim}
	\end{figure}

	
	Next, we vary the coupling phases to study their impact on laser synchronization. 
	Since only the relative phases matter, we fix the coupling phase $\phi_{23}$ between lasers 2 and 3, and scan $\phi_{12}$ between 1 and 2 and $\phi_{13}$ between 1 and 3. We compare uniform and proportional coupling for 31 combinations of $\phi_{12} - \phi_{23}$ and $\phi_{13} - \phi_{23}$. With uniform coupling, 10 out of 31 coupling phases cannot synchronize at the maximum coupling budget experimentally. The remaining 21 exhibit a large variation of synchronization threshold, as shown in Fig.~\ref{fig:TH_Sim}(a). For proportional coupling, only 2 out of 31 coupling phases fail to synchronize, and the rest display smaller fluctuations of synchronization threshold than uniform coupling. The median of synchronization threshold for proportional coupling is $\sim 1.5$ times lower than that of uniform coupling. 
	
	To verify the phase-dependent synchronization, we run numerical simulations based on the Lang--Kobayashi equations for three coupled VCSELs \cite{LangKobayashi1980, Ye2025OptimalSparse, VigneSM26}. We used a delay of $\tau = 1$~ns and linewidth enhancement factor of $\alpha = 2$.
	Figure~\ref{fig:TH_Sim}(b) shows the calculated synchronization threshold as a function of $(\phi_{12} - \phi_{23},\, \phi_{13} - \phi_{23})$ for uniform coupling of three VCSELs. The two-dimensional (2D) map shows segregated regions of low (lighter color) and high (darker color) coupling budget needed for synchronization. Some regions (black) have no synchronization up to maximum coupling budget. Stars represent experimental data for uniform coupling. 
	The black crosses mark no synchronization observed experimentally. They are located in the vicinity of dark regions from numerical simulation. These results confirm the strong dependence of laser synchronization threshold on the coupling phases. 
	
	As shown in Fig.~\ref{fig:TH_Sim}(c), proportional coupling has more extended regions of lower synchronization threshold than uniform coupling. Not only the median but also the variation of thresholds are reduced, consistent with our experimental data. The relative difference between experimental and numerical synchronization thresholds is 26\%. The reported synchronization behaviors are robust and reproducible at varying pumping levels and with different frequency detuning, as confirmed experimentally and numerically  \cite{VigneSM26}.
	
	
	
	To understand how coupling phases can modify synchronization threshold, we examine the phases of locked lasers. In the steady-state synchronization, the field of laser $n$ is $E_n(t)=A_n e^{i(2 \pi \nu_l t+\theta_n)}$, where $A_n$ is the field amplitude, $\theta_n$ is the locked phase, and $\nu_l$ is the locked frequency. The field coupled from laser $m$ to $n$ is $A_m e^{i(2 \pi \nu_l t+\theta_m)} \, |\kappa_{nm}| e^{i \phi_{nm}} \, e^{- i 2 \pi \nu_l \tau}$. The coupling is most efficient when the field from laser $m$ to $n$ is in phase with the field of laser $n$, otherwise destructive interference of will increase the coupling budget for synchronization. The phase matching condition is given by
	\begin{equation}
		\theta_n = \theta_m + \phi_{nm} - 2 \pi \nu_l \tau + 2 \pi q \, ,
		\label{eq:PhaseMatch}
	\end{equation}
	where $q$ is an integer.	
	Similarly for laser $m$, the phase matching condition is $\theta_m = \theta_n + \phi_{mn} - 2 \pi \nu_l\tau + 2 \pi q_1$, where $q_1$ is also an integer. 
	
\begin{figure}[ht!]
		\centering
		\includegraphics[width=\linewidth]{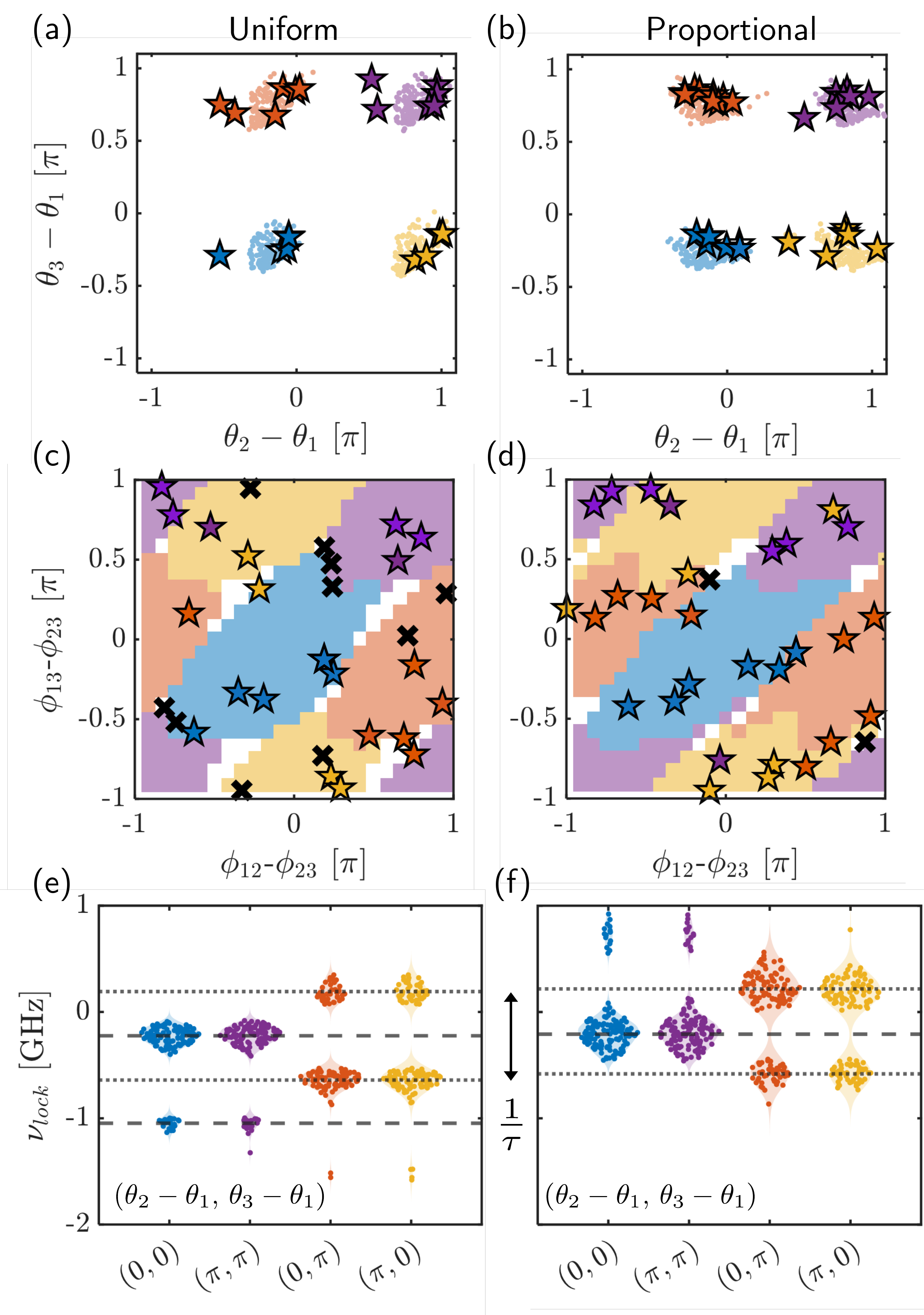}
		\caption{{\bf Locked laser frequency and relative phases.} Left column (a, c, d) for uniform coupling and right column (b, d, f) for proportional coupling of three VCSELs (identical to those in Fig.~2). (a, b) Relative phases of lasers 2 ($\theta_2$) and 3 ($\theta_3$) to 1 ($\theta_1$) at the synchronization threshold. Dots represent numerical results, and stars experimental data recovered from far-field interference patterns \cite{VigneSM26}. Colors of dots and inside stars encode the locked-state phases: blue for ($\theta_2-\theta_1$, $\theta_3-\theta_1$)~=~($0$, $0$), red for ($0$, $\pi$), yellow for ($\pi$, $0$) and purple for ($\pi$, $\pi$).        
        (c, d) Correspondence between relative phases of locked lasers and those of their coupling coefficients, from numerical simulation (colored regions) and experiment (stars and crosses). The color within each star represents the locked state phase and cross means no locking. (e, f) Locked frequencies at the synchronization threshold for different locked phases of three lasers (color). Four sets of locked phases $( \theta_2 - \theta_1, \theta_3 - \theta_1)$ give two sets of locked frequencies, which are equally spaced and offset by half the spacing. } 
		\label{fig:PhaseImpact}
	\end{figure}

	For symmetric reciprocal coupling, we have $ \phi_{nm} = \phi_{mn}$. Subtracting the reciprocal phase matching conditions leads to $ \theta_n - \theta_m = \pi (q-q_1)$. Thus the two lasers are locked either in phase $ \theta_n - \theta_m = 0$ or out of phase $ \theta_n - \theta_m = \pi $. This result holds for every pair of lasers satisfying the phase-matching condition. It is consistent with numerical results and experimental data for both uniform and proportional coupling in Figs~\ref{fig:PhaseImpact}(a, b). 

	When a laser is coupled with more than one lasers, the overall coupling is enhanced when the coupled fields from different lasers are in phase. The phase matching condition for fields from lasers $m$ and $l$ to $n$ is $\theta_m + \phi_{nm} - 2 \pi \nu_l\tau = \theta_l + \phi_{nl} - 2 \pi \nu_l\tau + 2 \pi q_2$, where $q_2$ is an integer. It gives the coupling phase difference $\phi_{nm} - \phi_{nl}  = \theta_l - \theta_m + 2 \pi q_2$.  The phase-locked lasers $l$ and $m$ have $\theta_l - \theta_m = 0, \pi$, thus $\phi_{nm} - \phi_{nl} = 0, \pi$, the coupling phase difference 
	compensate for the phase difference between locked lasers. The correlation between coupling phases and locked phases of lasers agrees with experimental and numerical results for both uniform and proportional coupling of three VCSELs in Figs~\ref{fig:PhaseImpact}(c, d).  
	
	However, phase-matching of coupled fields from different lasers to one laser does not guarantee they are in phase with the field of this laser. To ensure the field from laser $m$ to $n$ is in phase with the field of laser $n$, Eq.~\ref{eq:PhaseMatch} must be satisfied by adjusting the locked frequency $\nu_l$ to satisfy $ 2 \pi \nu_l \tau = \theta_m - \theta_n + \phi_{nm}$. For three coupled lasers with $\phi_{23}$ set to $0$, because $\theta_m - \theta_n = 0, \pi$ and $\phi_{nm} = 0, \pi$, we get $2 \pi \nu_l \tau = 0, \pi$. It gives two sets of locked frequencies. The first set is $\nu_l^{(1)} \tau = j$, and the second set $\nu_l^{(2)} \tau = j + 1/2$, where $j$ is an integer. The locked frequencies are equally spaced, but offset by half the spacing between the two sets. These are confirmed for three VCSELs with uniform and proportional coupling in Fig.~\ref{fig:PhaseImpact}(e, f).
	
	The phase-matching conditions ensure in-phase coupling, which fully utilizes a given coupling budget for laser synchronization. They result in regions of low synchronization thresholds in the coupled-phase maps for uniform and proportional coupling in Fig.~\ref{fig:PhaseImpact}(b,c). If phase-matching conditions are not satisfied, destructive interference of coupled fields will increase the coupling budget for synchronization. The phase frustration may be strong enough to block synchronization. For proportional coupling, the links between lasers have varying strength. The synchronization threshold is dictated by whether strongly-coupled fields are in phase. Even if weakly-coupled fields are not phase matched, their impact on synchronization is relatively small. Hence,  phase-matching conditions are relaxed for proportional coupling, and the regions of lower synchronization threshold are larger than uniform coupling with identical links for all laser pairs.

	
	Recent studies have shown that intentionally introducing disorder may facilitate synchronization of nonlinear oscillators and counter-balance the intrinsic disorder of laser frequencies \cite{Nair2021Disorder, barioni2025interpretable, OcampoEspindola2025Detuning}. However, the additional disorder is uncorrelated with the existing disorder in a laser array. Here we correlate the disorder introduced to laser coupling coefficient with the intrinsic frequency disorder. The `matched' disorder leads to an efficient allocation of a finite coupling budget among lasers with varying intrinsic frequencies. We further show that the coupling phases can be tuned to ensure constructive interference of coupled fields, further lowering the coupling budget for global synchronization. 
	
	
	While this work provides a proof of concept with three VCSELs, the proportional coupling scheme and phase-matching conditions are general and applicable to any number of lasers. 
	In the present study, symmetric coupling of lasers limits the relative phases of locked lasers to $0$ or $\pi$. An extension to asymmetric coupling will provide additional control over the locked state, allowing arbitrary phase difference between lasers. This will enable output beam steering and arbitrary pattern formation in the far-field. More broadly, independently-controlled non-reciprocal links will make it possible to engineer non-Hermitian and synthetic-gauge dynamics of nonlinear oscillator networks \cite{teimourpour2016non, Arwas2022AnyonicPT, hokmabadi2019supersymmetric, dikopoltsev2021topological, Gao2023GaugedLaserArray}.
	
	
	\begin{acknowledgments}
		This work was supported by the Office of Naval Research under Grant No. N00014-24-1-2548. N.V. and H.C. acknowledge useful discussions with Herbert G. Winful and Maxwell M Chumley of University of Michigan. 
	\end{acknowledgments}
	
	\bibliography{references}

@article{Fan2005,
  author={Fan, T.Y.},
  journal={IEEE J. Sel. Top. Quantum Electron.}, 
  title={Laser beam combining for high-power, high-radiance sources}, 
  year={2005},
  volume={11},
  number={3},
  pages={567-577},
  doi={10.1109/JSTQE.2005.850241}
}

@article{Kapon1984,
  author = {E. Kapon and J. Katz and A. Yariv},
  journal = {Opt. Lett.},
  number = {4},
  pages = {125--127},
  publisher = {Optica Publishing Group},
  title = {Supermode analysis of phase-locked arrays of semiconductor lasers},
  volume = {9},
  month = {Apr},
  year = {1984},
  url = {https://opg.optica.org/ol/abstract.cfm?URI=ol-9-4-125},
  doi = {10.1364/OL.9.000125},
}

@article{WinfulWang1988,
  author       = {Winful, H G and Wang, S S},
  title        = {Stability of phase locking in coupled semiconductor laser arrays},
  doi          = {10.1063/1.100363},
  url          = {https://www.osti.gov/biblio/6748013},
  journal      = {Appl. Phys. Lett.; (United States)},
  issn         = {ISSN APPLA},
  volume       = {53:20},
  place        = {United States},
  year         = {1988},
  month        = {11}
}

@article{Nair2018WeaklyCoupled,
  author = {Niketh Nair and Erik Bochove and Yehuda Braiman},
  journal = {Opt. Express},
  number = {16},
  pages = {20040--20050},
  publisher = {Optica Publishing Group},
  title = {Phase-locking of arrays of weakly coupled semiconductor lasers},
  volume = {26},
  month = {Aug},
  year = {2018},
  url = {https://opg.optica.org/oe/abstract.cfm?URI=oe-26-16-20040},
  doi = {10.1364/OE.26.020040},
}

@article{Argyris:16,
  author = {A. Argyris and M. Bourmpos and D. Syvridis},
  journal = {Opt. Express},
  number = {5},
  pages = {5600--5614},
  publisher = {Optica Publishing Group},
  title = {Experimental synchrony of semiconductor lasers in coupled networks},
  volume = {24},
  month = {Mar},
  year = {2016},
  url = {https://opg.optica.org/oe/abstract.cfm?URI=oe-24-5-5600},
  doi = {10.1364/OE.24.005600},
}

@ARTICLE{LangKobayashi1980,
       author = {{Lang}, R. and {Kobayashi}, K.},
        title = "{External optical feedback effects on semiconductor injection laser properties}",
      journal = {IEEE J. Quantum Electron.},
         year = 1980,
        month = mar,
       volume = {16},
       number = {3},
        pages = {347-355},
          doi = {10.1109/JQE.1980.1070479},
       adsurl = {https://ui.adsabs.harvard.edu/abs/1980IJQE...16..347L}
}

@ARTICLE{Henry1982,
       author = {{Henry}, C.},
        title = "{Theory of the linewidth of semiconductor lasers}",
      journal = {IEEE J. Quantum Electron.},
         year = 1982,
        month = feb,
       volume = {18},
       number = {2},
        pages = {259-264},
          doi = {10.1109/JQE.1982.1071522},
       adsurl = {https://ui.adsabs.harvard.edu/abs/1982IJQE...18..259H}
}

@article{Mulet2002Bidirectional,
  title = {Modeling bidirectionally coupled single-mode semiconductor lasers},
  author = {Mulet, Josep and Masoller, Cristina and Mirasso, Claudio R.},
  journal = {Phys. Rev. A},
  volume = {65},
  issue = {6},
  pages = {063815},
  numpages = {12},
  year = {2002},
  month = {Jun},
  publisher = {American Physical Society},
  doi = {10.1103/PhysRevA.65.063815},
  url = {https://link.aps.org/doi/10.1103/PhysRevA.65.063815}
}

@article{Erzgraber2005Bifurcation,
    title = {Mutually delay-coupled semiconductor lasers: Mode bifurcation scenarios},
    journal = {Opt. Commun.},
    volume = {255},
    number = {4},
    pages = {286-296},
    year = {2005},
    issn = {0030-4018},
    doi = {https://doi.org/10.1016/j.optcom.2005.06.016},
    url = {https://www.sciencedirect.com/science/article/pii/S003040180500564X},
    author = {H. Erzgräber and D. Lenstra and B. Krauskopf and E. Wille and M. Peil and I. Fischer and W. Elsäßer}
}

@article{Wunsche2005DelayCoupled,
  title = {Synchronization of Delay-Coupled Oscillators: A Study of Semiconductor Lasers},
  author = {W\"unsche, H.-J. and Bauer, S. and Kreissl, J. and Ushakov, O. and Korneyev, N. and Henneberger, F. and Wille, E. and Erzgr\"aber, H. and Peil, M. and Els\"a\ss{}er, W. and Fischer, I.},  journal = {Phys. Rev. Lett.},
  volume = {94},
  issue = {16},
  pages = {163901},
  numpages = {4},
  year = {2005},
  month = {Apr},
  publisher = {American Physical Society},
  doi = {10.1103/PhysRevLett.94.163901},
  url = {https://link.aps.org/doi/10.1103/PhysRevLett.94.163901}
}

@article{Soriano2013ComplexPhotonics,
  title = {Complex photonics: Dynamics and applications of delay-coupled semiconductors lasers},
  author = {Soriano, Miguel C. and Garc{\'i}a-Ojalvo, Jordi and Mirasso, Claudio R. and Fischer, Ingo},
  journal = {Rev. Mod. Phys.},
  volume = {85},
  issue = {1},
  pages = {421--470},
  numpages = {0},
  year = {2013},
  month = {Mar},
  publisher = {American Physical Society},
  doi = {10.1103/RevModPhys.85.421},
  url = {https://link.aps.org/doi/10.1103/RevModPhys.85.421}
}

@book{Ohtsubo2017,
  author    = {Ohtsubo, Junji},
  title     = {Semiconductor Lasers: Stability, Instability and Chaos},
  edition   = {4},
  series    = {Springer Series in Optical Sciences},
  volume    = {111},
  publisher = {Springer},
  year      = {2017},
  doi       = {10.1007/978-3-319-56138-7}
}

@article{Brunner:15,
  author = {Daniel Brunner and Ingo Fischer},
  journal = {Opt. Lett.},
  number = {16},
  pages = {3854--3857},
  publisher = {Optica Publishing Group},
  title = {Reconfigurable semiconductor laser networks based on diffractive coupling},
  volume = {40},
  month = {Aug},
  year = {2015},
  url = {https://opg.optica.org/ol/abstract.cfm?URI=ol-40-16-3854},
  doi = {10.1364/OL.40.003854},
}

@article{Arwas2022AnyonicPT,
  author = {Geva Arwas  and Sagie Gadasi  and Igor Gershenzon  and Asher Friesem  and Nir Davidson  and Oren Raz },
  title = {Anyonic-parity-time symmetry in complex-coupled lasers},
  journal = {Sci. Adv.},
  volume = {8},
  number = {22},
  pages = {eabm7454},
  year = {2022},
  doi = {10.1126/sciadv.abm7454},
  URL = {https://www.science.org/doi/abs/10.1126/sciadv.abm7454}
}

@article{Gao2023GaugedLaserArray,
  title = {Two-Dimensional Reconfigurable Non-Hermitian Gauged Laser Array},
  author = {Gao, Zihe and Qiao, Xingdu and Pan, Mingsen and Wu, Shuang and Yim, Jieun and Chen, Kaiyuan and Midya, Bikashkali and Ge, Li and Feng, Liang},
  journal = {Phys. Rev. Lett.},
  volume = {130},
  issue = {26},
  pages = {263801},
  numpages = {6},
  year = {2023},
  month = {Jun},
  publisher = {American Physical Society},
  doi = {10.1103/PhysRevLett.130.263801},
  url = {https://link.aps.org/doi/10.1103/PhysRevLett.130.263801}
}

@misc{Ye2025OptimalSparse,
      title={Optimal sparse networks for synchronization of semiconductor lasers}, 
      author={Li-Li Ye and Nathan Vigne and Fan-Yi Lin and Hui Cao and Ying-Cheng Lai},
      year={2025},
      eprint={2511.03205},
      archivePrefix={arXiv},
      primaryClass={physics.optics},
      url={https://arxiv.org/abs/2511.03205}, 
}

@misc{MikaberidzeTaylor2025Emergent,
      title={Emergent Topology of Optimal Networks for Synchrony}, 
      author={Guram Mikaberidze and Dane Taylor},
      year={2026},
      eprint={2509.18279},
      archivePrefix={arXiv},
      primaryClass={nlin.AO},
      url={https://arxiv.org/abs/2509.18279}, 
}

@misc{Pando2026Proportional,
      title={Enhanced synchronization with proportional coupling in Kuramoto oscillator networks}, 
      author={Amit Pando and Eran Bernstein and Tomer Hacohen and Nathan Vigne and Hui Cao and Oren Raz and Asher Friesem and Nir Davidson},
      year={2026},
      eprint={2603.29648},
      archivePrefix={arXiv},
      primaryClass={cond-mat.stat-mech},
      url={https://arxiv.org/abs/2603.29648}, 
}

@article{Nair2021Disorder,
  title = {Using Disorder to Overcome Disorder: A Mechanism for Frequency and Phase Synchronization of Diode Laser Arrays},
  author = {Nair, N. and Hu, K. and Berrill, M. and Wiesenfeld, K. and Braiman, Y.},
  journal = {Phys. Rev. Lett.},
  volume = {127},
  issue = {17},
  pages = {173901},
  numpages = {7},
  year = {2021},
  month = {Oct},
  publisher = {American Physical Society},
  doi = {10.1103/PhysRevLett.127.173901},
  url = {https://link.aps.org/doi/10.1103/PhysRevLett.127.173901}
}

@article{barioni2025interpretable,
  title={Interpretable disorder-promoted synchronization and coherence in coupled laser networks},
  author={Barioni, Ana Elisa D and Montanari, Arthur N and Motter, Adilson E},
  journal={Phys. Rev. Lett.},
  volume={135},
  number={19},
  pages={197401},
  year={2025},
  publisher={APS}
}

@article{OcampoEspindola2025Detuning,
author = {Jorge Luis Ocampo-Espindola  and Christian Bick  and Adilson E. Motter  and István Z. Kiss },
title = {Frequency synchronization induced by frequency detuning},
journal = {Sci. Adv.},
volume = {11},
number = {24},
pages = {eadu4114},
year = {2025},
doi = {10.1126/sciadv.adu4114},
URL = {https://www.science.org/doi/abs/10.1126/sciadv.adu4114}
}

@article{Nixon2013GeometricFrustration,
  title = {Observing Geometric Frustration with Thousands of Coupled Lasers},
  author = {Nixon, Micha and Ronen, Eitan and Friesem, Asher A. and Davidson, Nir},
  journal = {Phys. Rev. Lett.},
  volume = {110},
  issue = {18},
  pages = {184102},
  numpages = {5},
  year = {2013},
  month = {May},
  publisher = {American Physical Society},
  doi = {10.1103/PhysRevLett.110.184102},
  url = {https://link.aps.org/doi/10.1103/PhysRevLett.110.184102}
}

@article{Skalli2022VCSELNeuromorphic,
  author = {Anas Skalli and Joshua Robertson and Dafydd Owen-Newns and Matej Hejda and Xavier Porte and Stephan Reitzenstein and Antonio Hurtado and Daniel Brunner},
  journal = {Opt. Mater. Express},
  number = {6},
  pages = {2395--2414},
  publisher = {Optica Publishing Group},
  title = {Photonic neuromorphic computing using vertical cavity semiconductor lasers},
  volume = {12},
  month = {Jun},
  year = {2022},
  url = {https://opg.optica.org/ome/abstract.cfm?URI=ome-12-6-2395},
  doi = {10.1364/OME.450926},
}

@article{kozyreff2000global,
  title={Global coupling with time delay in an array of semiconductor lasers},
  author={Kozyreff, Gregory and Vladimirov, AG and Mandel, Paul},
  journal={Phys. Rev. Lett.},
  volume={85},
  number={18},
  pages={3809},
  year={2000},
  publisher={APS}
}

@article{zamora2010crowd,
  title={Crowd synchrony and quorum sensing in delay-coupled lasers},
  author={Zamora-Munt, Jordi and Masoller, C and Garcia-Ojalvo, Jordi and Roy, Rajarshi},
  journal={Phys. Rev. Lett.},
  volume={105},
  number={26},
  pages={264101},
  year={2010},
  publisher={APS}
}

@article{bourmpos2012sensitivity,
  title={Sensitivity analysis of a star optical network based on mutually coupled semiconductor lasers},
  author={Bourmpos, Michail and Argyris, Apostolos and Syvridis, Dimitris},
  journal={J. Lightwave Technol.},
  volume={30},
  number={16},
  pages={2618--2624},
  year={2012},
  publisher={OSA}
}

@article{xiang2016synchronization,
  title={Synchronization regime of star-type laser network with heterogeneous coupling delays},
  author={Xiang, ShuiYing and Wen, AiJun and Pan, Wei},
  journal={IEEE Photonics Technol. Lett.},
  volume={28},
  number={18},
  pages={1988--1991},
  year={2016},
  publisher={IEEE}
}

@article{nixon2012controlling,
  title={Controlling synchronization in large laser networks},
  author={Nixon, Micha and Fridman, Moti and Ronen, Eitan and Friesem, Asher A and Davidson, Nir and Kanter, Ido},
  journal={Phys. Rev. Lett.},
  volume={108},
  number={21},
  pages={214101},
  year={2012},
  publisher={APS}
}

@article{flunkert2012chaos,
  title={Chaos synchronization in networks of delay-coupled lasers: role of the coupling phases},
  author={Flunkert, Valentin and Sch\"oll, Eckehard},
  journal={New J. Phys.},
  volume={14},
  number={3},
  pages={033039},
  year={2012},
  publisher={IOP Publishing}
}

@article{Arenas2008SynchronizationComplexNetworks,
title = {Synchronization in complex networks},
journal = {Phys. Rep.},
volume = {469},
number = {3},
pages = {93-153},
year = {2008},
issn = {0370-1573},
doi = {https://doi.org/10.1016/j.physrep.2008.09.002},
url = {https://www.sciencedirect.com/science/article/pii/S0370157308003384},
author = {Alex Arenas and Albert Díaz-Guilera and Jurgen Kurths and Yamir Moreno and Changsong Zhou}
}

@article{chen2023deep,
  title={Deep learning with coherent {VCSEL} neural networks},
  author={Chen, Zaijun and Sludds, Alexander and Davis, Ronald and Christen, Ian and Bernstein, Liane and Heuser, Tobias and Heermeier, Niels and Lott, James A and Reitzenstein, Stephan and Hamerly, Ryan and others},
  journal={Nat. Photonics},
  volume={17},
  number={8},
  pages={723--730},
  year={2023},
  publisher={Nature Publishing Group UK London}
}

@article{LiuBraiman2013CBC,
author = {B. Liu and Y. Braiman},
journal = {Opt. Express},
number = {25},
pages = {31218--31228},
publisher = {Optica Publishing Group},
title = {Coherent beam combining of high power broad-area laser diode array with near diffraction limited beam quality and high power conversion efficiency},
volume = {21},
month = {Dec},
year = {2013},
url = {https://opg.optica.org/oe/abstract.cfm?URI=oe-21-25-31218},
doi = {10.1364/OE.21.031218},
}

@article{orenstein1992Large2DVCSEL,
    author = {Orenstein, M. and Kapon, E. and Harbison, J. P. and Florez, L. T. and Stoffel, N. G.},
    title = {Large two‐dimensional arrays of phase‐locked vertical cavity surface emitting lasers},
    journal = {Appl. Phys. Lett.},
    volume = {60},
    number = {13},
    pages = {1535-1537},
    year = {1992},
    month = {03},
    issn = {0003-6951},
    doi = {10.1063/1.107243},
    url = {https://doi.org/10.1063/1.107243},
}

@article{SakaguchiKuramoto1986,
    author = {Sakaguchi, Hidetsugu and Kuramoto, Yoshiki},
    title = {A Soluble Active Rotator Model Showing Phase Transitions via Mutual Entrainment},
    journal = {Prog. Theor. Phys.},
    volume = {76},
    number = {3},
    pages = {576-581},
    year = {1986},
    month = {09},
    issn = {0033-068X},
    doi = {10.1143/PTP.76.576},
    url = {https://doi.org/10.1143/PTP.76.576},
}

@article{lei2023new,
  title={A new criterion for optimizing synchrony of coupled oscillators},
  author={Lei, Yong and Xu, Xin-Jian and Wang, Xiaofan and Zou, Yong and Kurths, J{\"u}rgen},
  journal={Chaos, Solitons \& Fractals},
  volume={168},
  pages={113192},
  year={2023},
  publisher={Elsevier}
}

@article{skardal2014optimal,
  title={Optimal synchronization of complex networks},
  author={Skardal, Per Sebastian and Taylor, Dane and Sun, Jie},
  journal={Phys. Rev. Lett.},
  volume={113},
  number={14},
  pages={144101},
  year={2014},
  publisher={APS}
}

@article{brede2008synchrony,
  title={Synchrony-optimized networks of non-identical Kuramoto oscillators},
  author={Brede, Markus},
  journal={Phys. Lett. A},
  volume={372},
  number={15},
  pages={2618--2622},
  year={2008},
  publisher={Elsevier}
}

@article{kelly2011topology,
  title={On the topology of synchrony optimized networks of a Kuramoto-model with non-identical oscillators},
  author={Kelly, David and Gottwald, Georg A},
  journal={Chaos},
  volume={21},
  number={2},
  year={2011},
  publisher={AIP Publishing}
}

@article{arenas2008synchronization,
  title={Synchronization in complex networks},
  author={Arenas, Alex and D{\'\i}az-Guilera, Albert and Kurths, Jurgen and Moreno, Yamir and Zhou, Changsong},
  journal={Phys. Rep.},
  volume={469},
  number={3},
  pages={93--153},
  year={2008},
  publisher={Elsevier}
}

@article{dorogovtsev2008critical,
  title={Critical phenomena in complex networks},
  author={Dorogovtsev, Sergey N and Goltsev, Alexander V and Mendes, Jos{\'e} FF},
  journal={Rev. Mod. Phys.},
  volume={80},
  number={4},
  pages={1275--1335},
  year={2008},
  publisher={APS}
}

@article{liu2004synchronization,
  title={Synchronization of high-power broad-area semiconductor lasers},
  author={Liu, Yun and Braiman, Yehuda},
  journal={IEEE J. Sel. Top. Quantum Electron.},
  volume={10},
  number={5},
  pages={1013--1024},
  year={2004},
  publisher={IEEE}
}

@article{hokmabadi2019supersymmetric,
  title={Supersymmetric laser arrays},
  author={Hokmabadi, Mohammad P and Nye, Nicholas S and El-Ganainy, Ramy and Christodoulides, Demetrios N and Khajavikhan, Mercedeh},
  journal={Science},
  volume={363},
  number={6427},
  pages={623--626},
  year={2019},
  publisher={American Association for the Advancement of Science}
}

@article{teimourpour2016non,
  title={Non-Hermitian engineering of single mode two dimensional laser arrays},
  author={Teimourpour, Mohammad H and Ge, Li and Christodoulides, Demetrios N and El-Ganainy, Ramy},
  journal={Scientific reports},
  volume={6},
  number={1},
  pages={33253},
  year={2016},
  publisher={Nature Publishing Group UK London}
}

@article{dikopoltsev2021topological,
  title={Topological insulator vertical-cavity laser array},
  author={Dikopoltsev, Alex and Harder, Tristan H and Lustig, Eran and Egorov, Oleg A and Beierlein, Johannes and Wolf, Adriana and Lumer, Yaakov and Emmerling, Monika and Schneider, Christian and H{\"o}fling, Sven and others},
  journal={Science},
  volume={373},
  number={6562},
  pages={1514--1517},
  year={2021},
  publisher={American Association for the Advancement of Science}
}

@article{nixon2011synchronized,
  title={Synchronized cluster formation in coupled laser networks},
  author={Nixon, Micha and Friedman, Moti and Ronen, Eitan and Friesem, Asher A and Davidson, Nir and Kanter, Ido},
  journal={Phys. Rev. Lett.},
  volume={106},
  number={22},
  pages={223901},
  year={2011},
  publisher={APS}
}

@misc{VigneSM26,
    key = {},
    note = {See Supplemental Material at [URL will be inserted by publisher] for more detailed experimental, numerical methods and additional results},
}

@article{nixon2009PhaseLocking,
author = {Micha Nixon and Moti Fridman and Eitan Ronen and Asher A. Friesem and Nir Davidson},
journal = {Opt. Lett.},
number = {12},
pages = {1864--1866},
publisher = {Optica Publishing Group},
title = {Phase locking of two fiber lasers with time-delayed coupling},
volume = {34},
month = {Jun},
year = {2009},
url = {https://opg.optica.org/ol/abstract.cfm?URI=ol-34-12-1864},
doi = {10.1364/OL.34.001864},
}

@ARTICLE{adler1946Injection,
  author={Adler, R.},
  journal={Proc. IRE}, 
  title={A Study of Locking Phenomena in Oscillators}, 
  year={1946},
  volume={34},
  number={6},
  pages={351-357},
  doi={10.1109/JRPROC.1946.229930}}

	\clearpage

	\begin{widetext}
	\begin{center}
	\textbf{\large Supplementary material:\\ 
		Tailoring complex coupling for efficient synchronization of semiconductor lasers}
	\end{center}
	\end{widetext}

	\setcounter{equation}{0}
	\setcounter{figure}{0}
	\setcounter{table}{0}
	\setcounter{page}{1}
	\makeatletter
	\renewcommand{\theequation}{S\arabic{equation}}
	\renewcommand{\thefigure}{S\arabic{figure}}
	\renewcommand{\bibnumfmt}[1]{[S#1]}
	\renewcommand{\citenumfont}[1]{S#1}

	\section{Laser characterization}
	
\begin{figure}[ht]
		\centering
		\includegraphics[width=\linewidth]{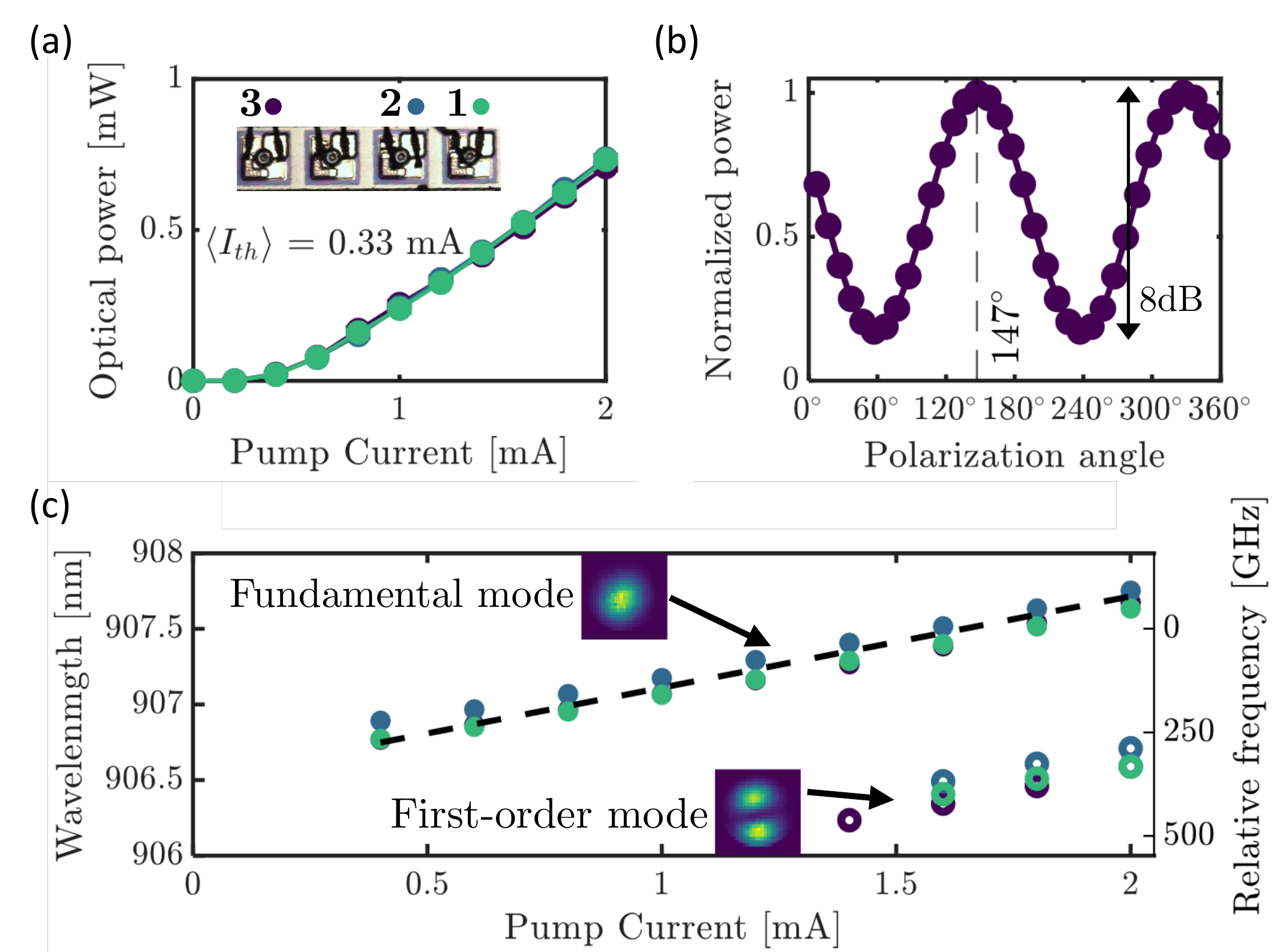}
		\caption{{\bf VCSEL array characterization}. (a) Emitted optical power as a function of pump current for three VCSELs (labeled 1, 2, 3) in a one-dimensional array, which correspond to Fig. 2 in the main text. Inset: a photo of the VCSEL array showing lasers 1, 2, 3 with uneven spacing. The average lasing threshold current is \qty{0.33}{\milli\ampere}. (b) Normalized power of laser $3$ as a function of the polarization angle at pump current $I=\qty{1}{\milli\ampere}$. The polarization extinction ratio PER = 8 dB. (c) Lasing wavelength as a function of pump current for three VCSELs. Lasing occurs only in the fundamental transverse mode (solid circles) when the pump current is below 1.4 mA. Its wavelength increases linearly with pump current, the dashed line is a linear fit. Above 1.4 mA, lasing also occurs in the first-order transverse mode (open circles) at shorter wavelengths. Inset: far-field intensity images of the fundamental and first-order lasing modes.}
		\label{fig:SI_lasercarac}
	\end{figure}

	A one-dimensional (1D) vertical-cavity surface-emitting laser (VCSEL) array from VI-system \textit{GmbH} (VM100-910-SG-qSM) consists of 12 lasers separated by 250 $\mu$m. The inset of Fig.~\ref{fig:SI_lasercarac}(a) marks three VCSELs (labeled 1, 2, 3) used for the synchronization experiment in Fig.~2 of the main text. The electric currents injected to individual lasers are provided by separate current drivers. Emitted powers from lasers 1, 2, and 3 are plotted as a function of pump current in Fig.~\ref{fig:SI_lasercarac}(a). They have nearly identical powers and lasing thresholds. The mean threshold current is $\qty{0.33}{\milli\ampere}$. 
	
	We place a linear polarizer in front of one laser and measure the transmitted power as we rotate the linear polarizer. Figure~\ref{fig:SI_lasercarac}(b) shows power variation with polarization angle at the pump current of 1 mA for laser 3. The laser emission is linearly polarized in the direction of \qty{147}{\degree} from the horizontal plane. The ratio of maximum and minimum power gives the polarization extinction ratio PER~=~8~dB. Similar polarization characteristics are observed for other VCSELs.   
	
	Figure~\ref{fig:SI_lasercarac}(c) shows the lasing wavelength as a function of pump current. Below 1.4 mA, lasing occurs only in the fundamental transverse mode at wavelength around \qty{907}{\nano\meter}. The spectrally-resolved far-field image shows the emission intensity distribution of the fundamental mode in the inset of Fig.~\ref{fig:SI_lasercarac}(c).  Its wavelength $\lambda$ increases almost linearly with pump current $I$. A linear fit (dashed line) gives the slope  $d\lambda/dI = \qty{0.605}{\nano\meter/\milli\ampere}$. At $\lambda = \qty{907.5}{\nano\meter}$, this corresponds to a frequency change of \qty{-220}{\giga\hertz/\milli\ampere}.
	In the synchronization experiment, we set the pump current to 1 mA so that only the fundamental mode lases in each VCSEL. 
	
	The variation of lasing wavelength with pump current allows us to fine-tune the frequency difference between VCSELs. We use a high-precision current driver (Thorlabs LDC200CV) and battery-based low-noise drivers for precise control of currents injected to individual VCSELs. A 100 $\mu$A change in pump current led to a 22 GHz shift of optical frequency. Free-running frequency difference among VCSEL 1, 2 and 3 is within $\pm$ 20 GHz. Thus, small variations of their currents are sufficient to tune their frequency differences, without a notable change in their powers. 
	
	\begin{figure*}[ht!]
		\centering
		\includegraphics[width=\linewidth]{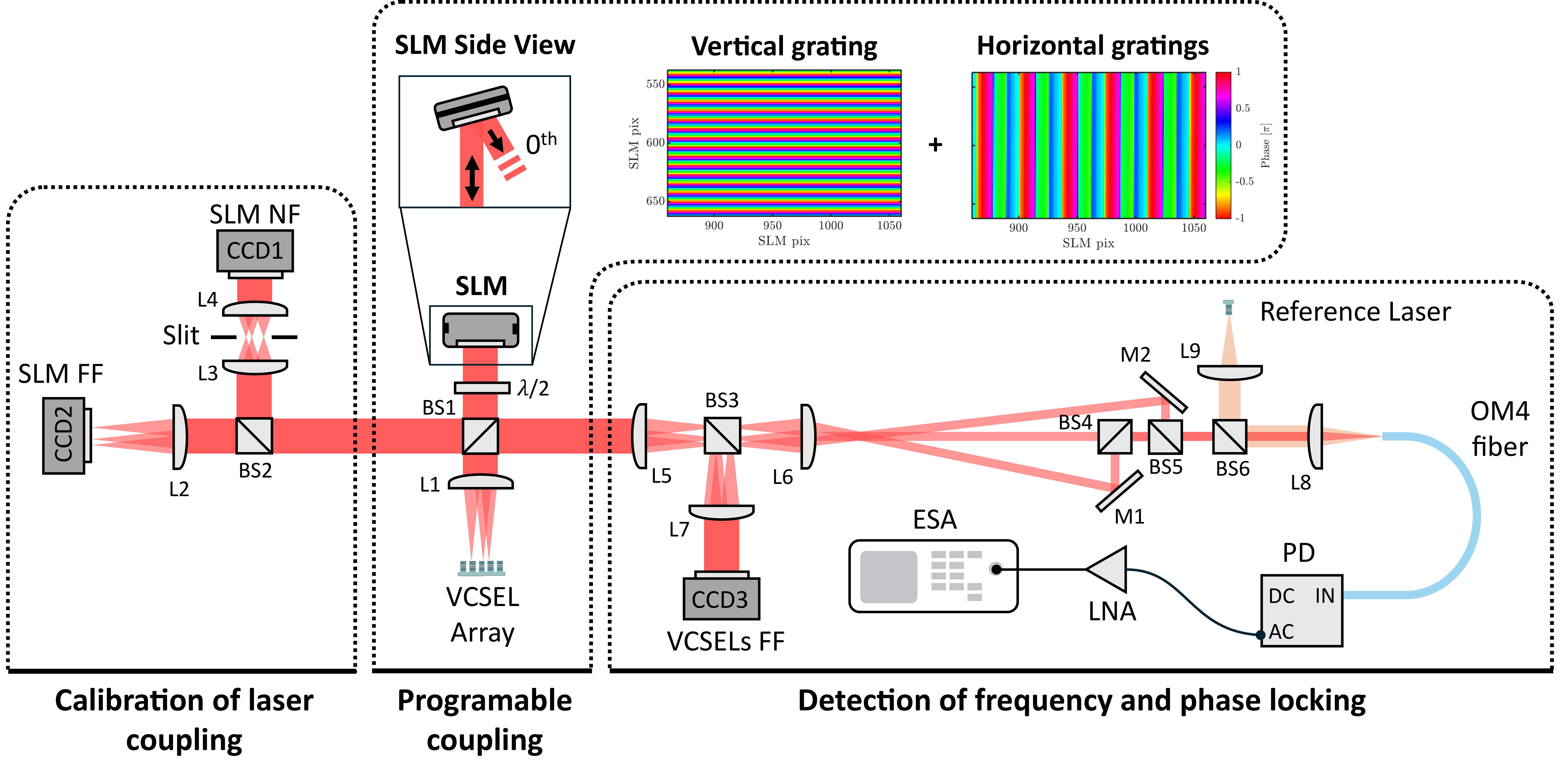}
		\caption{{\bf Schematic of experimental setup.} The entire setup consists of three modules: programmable coupling of lasers, calibration of laser coupling, and detection of frequency and phase locking. See text for a detailed description of each module.}
		\label{fig:SI_setup}
	\end{figure*}
	
	\section{Laser coupling setup}
	
	As shown schematically in Fig.~\ref{fig:SI_setup}, the experimental setup includes three modules: programmable coupling of lasers, calibration of laser coupling, detection of frequency and phase locking.  
	
	\subsection{Programmable coupling of lasers}
	
	A VCSEL array chip is positioned at the front focal plane of a lens (L1, focal length $f_1$ = \qty{80}{\milli\meter}), and a liquid-crystal-based phase-only spatial light modulator (SLM, Santec SLM-210) is placed at the back focal plane. A half-wave plate ($\lambda/2$) rotates the polarization direction of VCSEL emission to align with the SLM operating polarization. A 50:50 beam splitter (BS1) directs half of the laser emission to the detection module, and half of the reflection from the SLM to the calibration module. 
	
	The first-order diffraction of the SLM is used for laser coupling. A slight vertical tilt of the SLM directs the reflected beam ($0^{\mathrm{th}}$-order diffraction) away from the laser array, and a blazed grating vertically imprinted onto the SLM returns the first-order diffracted beam to the VCSEL array. The modulation depth of this phase grating is varied from 0 to $2 \pi$ to control the amount of diffracted power. 
	In addition to the vertical grating, a series of horizontal gratings of varying periods are written to the SLM to diffract emission from one laser to others in an 1D horizontal array.
	
	\subsection{Calibration of laser coupling}
	
	To monitor the laser emission incident onto the SLM, we take the near-field image of the SLM plane (SLM NF) using a pair of lenses (L3, L4) and a camera (CCD1, Allied Vision Mako G-125B) after a beam splitter (BS2). A slit in between the lenses filters the 0th-order diffraction. To calibrate the diffracted light back to the VCSEL array, we simultaneously take the far-field image of light diffracted from the SLM (SLM FF), which coincides with the VCSEL plane, using a lens (L2) and a camera (CCD2, Ximea MQ01RG-ON-S7 camera).     
	
	From the SLM FF image, we measure the coupling of lasers. We turn on only one laser, e.g., laser $m$, and take the SLM FF image without horizontal gratings on the SLM. It gives the total diffracted power $P_m$ for laser $m$. Then horizontal gratings are written to the SLM, and we measure the power of diffracted light $P_{nm}$ at laser $n$. The diffraction coefficient $R_{nm} = P_{nm} / P_m$ gives the normalized coupling amplitude $|\kappa_{nm}^{(e)}| = \sqrt{R_{nm}}$. The experimental coupling budget is $b_{\mathrm{tot}}^{\mathrm{exp}} = (1/M)\sum_{n,m} |\kappa_{nm}^{(e)}|$. 
	
	From the SLM NF image, we can extract the phase of the coupling coefficient using off-axis holography techniques. Since the diffracted beams from one laser to others are spatially extended in the SLM NF plane, it is more accurate to measure their phases on this plane than the SLM FF plane. We turn on only one laser at a time and redirect emission from the detection arm towards the SLM NF camera to serve as the reference beam $R$. For example, laser $m$ is turned on, its diffracted field $E_m = \sum_n E_{nm} = \sum_n \sqrt{R_{nm}} \, e^{i \phi_{nm}}$ interferes with $R$ to create an intensity pattern at the SLM NF camera,  $I_{\mathrm{SLMNF}} = |E_m|^2 + |R|^2 + E_m R^* + E_m^*R$, From the cross term $E_m R^*$, we retrieve the coupling phases $\phi_{nm}$.
	
	\subsection{Detection of frequency and phase locking}
	
	To detect frequency locking, we couple emission from VCSELs 1, 2, and 3 in the array to an optical fiber (OM4, \qty{50}{\micro\meter} core) using a series of lenses (L5-L8), mirrors (M1, M2) and beam splitters (BS4, BS5). The emission from a reference laser in another VCSEL chip is also coupled to the fiber with a lens (L9) and beam splitter (BS6). This fiber is connected to a  bandwidth-amplified photodetector (PD, New-Focus Model 1544 \qty{12}{\giga\hertz}). The electrical AC signal is further amplified by a low-noise amplifier (LNA, AT-LNA-0018-3825H, 50 MHz -- 20 GHz) and then measured by an electrical spectrum analyzer (ESA, N9030B-PXARF). The detected signals include beatings among three VCSELs and their beatings with the reference. To separate them, the reference laser is frequency detuned from three VCSELs by $\sim 8$ GHz. From their beating frequencies with the reference, we extract the relative frequencies of three VCSELs. The linewidth resolution is limited by the reference laser to about 50~MHz full-width-half-maximum.
	
	To detect phase locking, we take the far-field image of VCSEL emission (VCSEL FF) using a beam splitter (BS3), a lens (L7) and a camera (CCD3,  Allied Vision Prosilica GC660N). A horizontal slice of vertical fringes gives the fringe contrast $C = \langle (I_{\mathrm{max}}-I_{\mathrm{min}})/(I_{\mathrm{max}}+I_{\mathrm{min}}) \rangle$, averaged spatially.

	\section{Complex coupling coefficients}
	
	\begin{figure}[ht!]
		\centering
		\includegraphics[width=\linewidth]{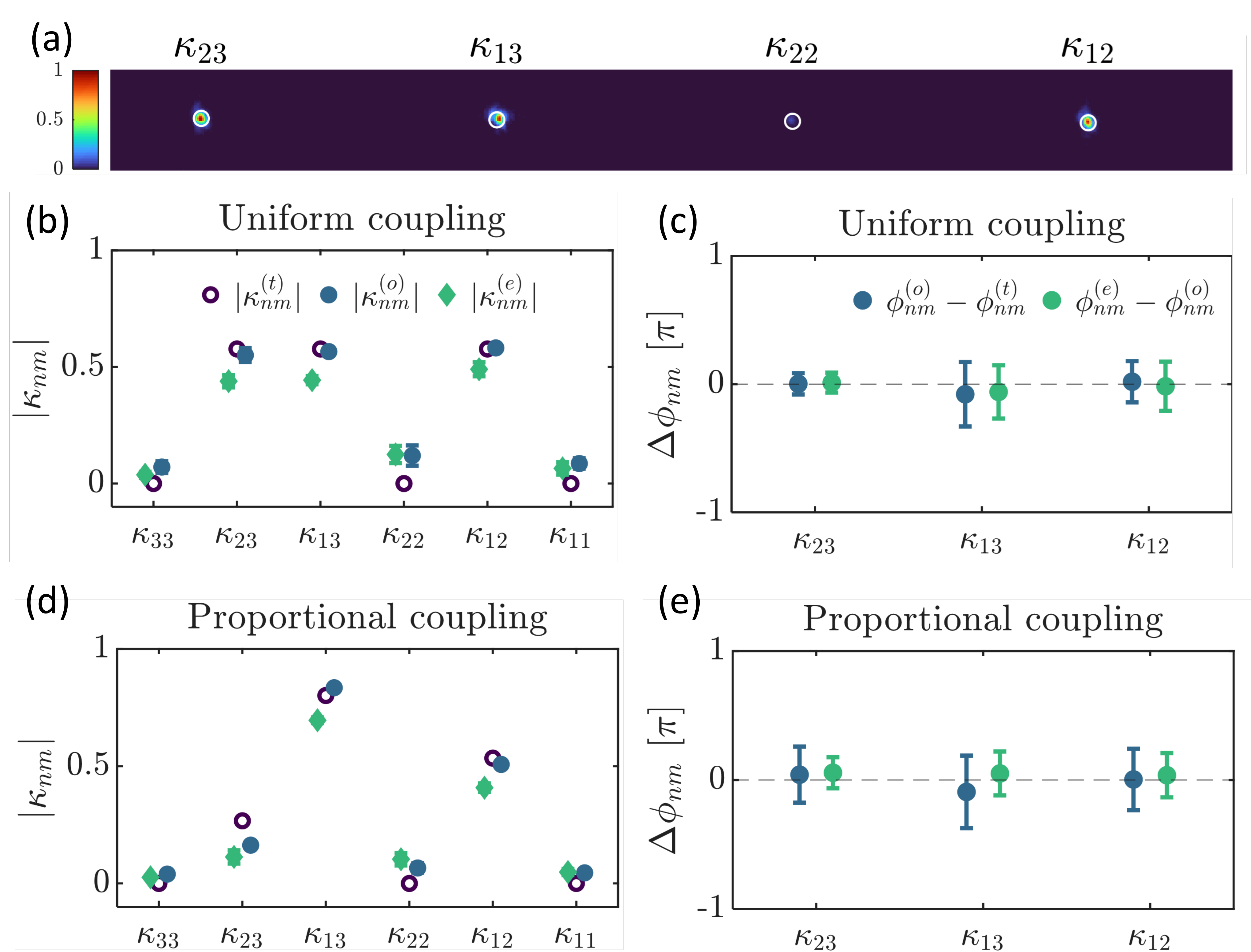}
		\caption{{\bf Accuracy of programming complex coupling coefficients.} (a) SLM farfield image (SLM FF) showing the diffracted beams from laser 2. Intensity $I$ is normalized to 1. White circles mark the locations of coupling coefficients.  (b, d) Normalized coupling amplitude $\kappa_{nm}^{(e)}$ measured experimentally (green) compared to target (purple) and numerically optimized (blue) values for uniform (b) and proportional (d) coupling. The error bar is the standard deviation over 31 different coupling phases. (c, e) Standard deviation of optimized coupling phase $\phi^{(\mathrm{o})}_{nm}$ from target $\phi^{(\mathrm{t})}_{nm}$ and measured coupling phase $\phi^{(\mathrm{e})}_{nm}$ from target $\phi^{(\mathrm{o})}_{nm}$ for uniform (c) and proportional (e) coupling. The measured phase was retrieved from off-axis holography measurement.}
		\label{fig:SI_patterns}
	\end{figure}
	
	For three VCSELs, we program independently three complex coupling coefficients: $\kappa_{12}$, $\kappa_{23}$, and $\kappa_{13}$. Meanwhile, we suppress self-coupling by keeping $\kappa_{11}$, $\kappa_{22}$, and $\kappa_{33}$ close to 0. With incident light from laser $m$, the diffracted light from the SLM generates multiple beams at the locations of other lasers. The field distribution at the VCSEL array (far-field of SLM) is given by $E_m = \sum_n \kappa_{nm} \, f(x_n)$, where $f(x_n)$ represents a Gaussian beam with a beam waist and divergence angle matching the VCSEL size and numerical aperture (NA) at the location $x_n$ of laser $n$. $x_n$ is the transverse coordinate with zero at the optical axis. For the three VCSELs, $x_1 = d$, $x_2$ = 0, $x_3 = -2d$, where $d = 250$ $\mu$m is the spacing between adjacent VCSELs.  
	
	Since the SLM is placed at the Fourier plane of the VCSEL array, the coupling coefficient $\kappa_{nm}$ depends solely on $\left(x_n + x_m \right)$. As an example, let us consider laser 2 that is on the optical axis $x_2 = 0$. If its distance to laser 1 on one side ($x_1 = d$) were the same as that to laser 3 on the other side ($x_3 = -d$), then $\kappa_{22} = \kappa_{13}$ since $\left(x_2+x_2\right) = \left(x_1 + x_3\right)$. To independently vary the coupling coefficients, the three lasers must be unevenly spaced. That is why we skip the laser at $x=-d$, in between 2 and 3 in the array [inset of Fig. S1(a)]. In the current selection of lasers, the SLM grating that couples lasers 1 and 3 also diffracts laser 2 emission to $x= -d$, the midpoint between 2 and 3. By measuring laser 2 emission diffracted to $x= -d$, we can calibrate $\kappa_{13}$. 
	
	We optimize the SLM phase pattern $\varphi_{\mathrm{SLM}}$ to generate the target field $E_m$ for a chosen set of complex coupling coefficients. First we build a forward model to compute $E_m$ from $\varphi_{\mathrm{SLM}}$, considering the finite SLM aperture and the VCSEL NA and system aberration. Next we construct a loss function to account for the difference between computed and targeted $E_m$ and add a penalty for power loss of coupled fields. For a target set of coupling coefficients $\kappa^{(t)}_{nm}$,  we numerically minimize the loss function using gradient-based optimization. The SLM phase pattern is updated using an Adam optimizer. The optimized coupling coefficients $\kappa^{(o)}_{nm}$ may not be exactly equal to the target $\kappa^{(t)}_{nm}$. Finally, the optimized SLM phase pattern is written to the SLM, and coupling coefficients are measured experimentally with the calibration module to ensure they are close to target values. 
	
	As an example, the measured uniform coupling pattern with laser $m=2$ on is shown in Fig.~\ref{fig:SI_patterns}(a). The image corresponds to the SLM far-field (SLM FF), which is equivalent to the VCSEL plane. White circles mark the positions corresponding to coupling terms $\kappa_{12}$, $\kappa_{13}$, $\kappa_{23}$ and self-coupling $\kappa_{22}$. From the power within each circle, we compute the experimental coupling amplitude $|\kappa_{nm}^{(e)}| = \sqrt{P^{(e)}_{nm}/P^{(e)}_m}$. Similarly, we compute the numerically optimized coupling amplitude $|\kappa_{nm}^{(o)}| = \sqrt{P^{(o)}_{nm}/P^{(o)}_m}$. We compare both experimental and numerically optimized values to the target values that are normalized with $|\kappa_{12}|^2+|\kappa_{13}|^2+|\kappa_{23}|^2 = 1$ in Fig.~\ref{fig:SI_patterns}(b) and (d) for uniform and proportional coupling. The measured $\kappa_{nm}^{(e)}$ shows a slightly lower value for $n \neq m$ and higher value for $n=m$ due to finite SLM diffraction efficiency and experimental inaccuracy.
	
	Coupling phases are retrieved from the off-axis holography measurement as described in the calibration module section. By sequentially turning on each laser, the coupling phases are retrieved for all lasers. Figures~\ref{fig:SI_patterns}(c) and (e) show the standard deviation of experimentally measured phases $\phi^{(\mathrm{e})}_{nm}$ averaged over all lasers from numerically optimized phases $\phi^{(\mathrm{o})}_{nm}$ in green and the standard deviation of $\phi^{(\mathrm{o})}_{nm}$ from target coupling phases $\phi^{(\mathrm{t})}_{nm}$ in blue for uniform and proportional coupling schemes. 
	
	\section{Coupling phase offset}
	
	Experimentally, the coupling phase $\phi_{nm}$ includes not only the phase shift imposed by the SLM ($\phi_{nm}^{\mathrm{(s)}}$) but also phases induced by time-delay $-2 \pi \nu_l \tau_{nm}$ and other optical misalignment. Small differences in diffraction path-lengths between individual laser pairs cause slight variations in delay times $\tau_{nm}$, which appear as static offsets of the pairwise coupling phases. Thus $\phi_{nm} = \phi_{nm}^{\mathrm{(s)}} + \phi_{nm}^{\mathrm{(\tau)}}$, where $\phi_{nm}^{\mathrm{(\tau)}} = 2 \pi \nu_l (\tau_{nm} - \tau)$, with $\tau$ the mean delay time. 
	Since we only measured the SLM-induced phase $\phi_{nm}^{\mathrm{(s)}}$, we fit the values of $\phi_{12}^{\mathrm{(\tau)}} - \phi_{23}^{\mathrm{(\tau)}}$ and $ \phi_{13}^{\mathrm{(\tau)}} - \phi_{23}^{\mathrm{(\tau)}}$ by maximizing the Pearson correlation between experimental and numerical synchronization threshold maps in Fig.~3(b,c) of the main text. Phase offsets to $\phi_{12} - \phi_{23}$ and $ \phi_{13} - \phi_{23}$ are identical for all data points obtained with uniform and proportional coupling. 
	
	\section{Laser phase extraction}	
	
\begin{figure}[ht!]
		\centering
		\includegraphics[width=\linewidth]{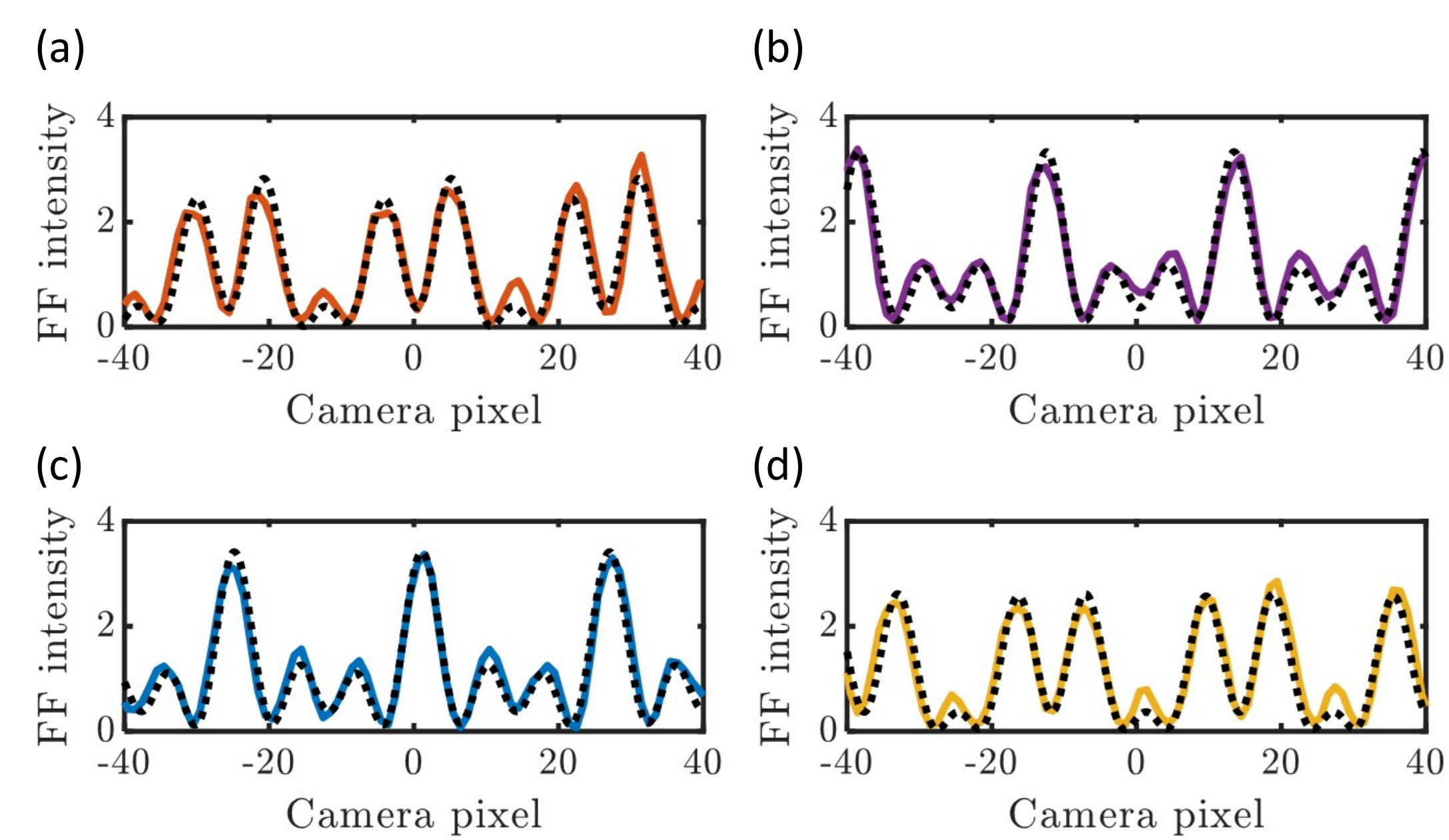}
		\caption{{\bf Far-field intensity fit to extract locked laser phases} (a-d) Experimentally measured (solid line) far-field (FF) intensity distribution and its fit (dashed line) to extract relative phases of three locked lasers. The fitted values of relative phases $(\theta_2-\theta_1,\,\theta_3-\theta_1)$ are (-0.094$\pi$, 0.86$\pi$) in (a), (0.97$\pi$, 0.88$\pi$) in (b), (-0.10$\pi$, -0.26$\pi$) in (c) and (0.89$\pi$,-0.29$\pi$) in (d).}
		\label{fig:SI_ffcalibration}
	\end{figure}

	Phase locking means that the phase difference between any two lasers is time-invariant, but not necessarily equal to 0. To extract relative phases of three VCSELs once they are locked, we fit the far-field interference pattern measured experimentally. For a horizontal array of lasers at locations $x_n$, the far-field intensity distribution is
	\begin{equation}
		I_{\mathrm{FF}}(k_x) = \left| \sum_{n=1}^{3} A_n e^{i\theta_n} e^{ -i k_x x_n} \right|^2
	\end{equation}
	where $A_n$ and $\theta_n$ represent the field amplitude and phase of laser $n$, $x_n$ is its transverse coordinate and $k_x$ is the transverse wavevector. Expanding this expression gives
	\begin{eqnarray}
		&&I_{\mathrm{FF}}(k_x) = \sum_{n=1}^{3} A_n^2 +\nonumber\\ 
		&&2 \sum_{n<m} A_n A_m \cos\left[k_x(x_n-x_m) - (\theta_n-\theta_m) \right]
	\end{eqnarray}
	By fitting $I_{\mathrm{FF}}(k_x)$, we extract the relative phase $\theta_n-\theta_m$ of lasers. We also include a small offset of the far-field center position, since it is difficult to pinpoint it experimentally. A few examples of far-field intensity fit to extract $\theta_2-\theta_1$ and $\theta_3-\theta_1$ for three VCSELs are shown in Fig.~\ref{fig:SI_ffcalibration}. 
	
	\section{Numerical simulations}
	
	The numerical results presented in the main text are obtained by solving the Lang-Kobayashi equations for three VCSELs.
	
	\begin{eqnarray}
		\frac{dE_n}{dt} =&& \frac{1+i\alpha}{2} \left( \frac{g_0\left(N_n-N_{\mathrm{tr}}\right)} {1+\epsilon |E_n(t)|^2} -\frac{1}{\tau_p} \right) E_n(t) \nonumber\\
		&&+i 2 \pi(\nu_n - \nu_0) E_n(t)
		\nonumber\\
		&&+ e^{-i 2 \pi \nu_0 \tau} \sum_{m \neq n}^{3} |\kappa_{nm}|e^{i\phi_{nm}}\,E_m(t-\tau) \nonumber\\
		&&+ F_{E_n}
		\label{eq:LK_field}
	\end{eqnarray}
	
	\begin{equation}
		\frac{dN_n}{dt} = J -\frac{N_n(t)}{\tau_s} - \frac{g_0\left(N_n-N_{\mathrm{tr}}\right)}{1+\epsilon |E_n(t)|^2} |E_n(t)|^2 + F_{N_n}
		\label{eq:LK_carrier}
	\end{equation}
	
	$E_n(t)$ is the slowly-varying complex electric field of laser $n$, $N_n(t)$ is its carrier density. $\alpha=2$ is the linewidth-enhancement factor that causes amplitude-phase coupling, $\tau_p = \qty{5.2}{\pico\second}$ and $\tau_s = \qty{0.25}{\nano\second}$ are the photon and carrier lifetimes, respectively. The carrier injection rate is  $J=3J_{th}$, where $J_{th}$ corresponds to the lasing threshold. $g_0 = \qty{8.75e-4}{\per\nano\second}$ represents the small-signal gain, $N_{tr} = 2.86\times10^5$ the carrier density at optical transparency, $\epsilon = 4\times10^{-6}$ the gain saturation factor.  $\nu_n$ is the free-running frequency of laser $n$, and $\nu_0$ is the mean frequency of all lasers. The complex coupling coefficient from laser $m$ to $n$ is given by $\kappa_{nm} = |\kappa_{nm}|e^{i\phi_{nm}}$. $\tau=\qty{1}{\nano\second}$ is the delay time in mutual coupling and is assumed constant for all laser pairs.  Stochastic noise for electric field is added through the Langevin term $\langle F_{E_n}(t) F^*_{E_n}(t^\prime)\rangle = (\beta_{sp}N_n / \tau_s) \, \delta(t-t^\prime)$ with $\beta_{sp} = 1.7\times10^{-3}$ the spontaneous emission factor, and for carrier density $\langle F_{N_n}(t) F_{N_n}(t^\prime)\rangle = (N_n / \tau_s) \, \delta(t-t^\prime)$ .
	
	To evaluate time-delayed coupling terms, fields at earlier times are stored over one delay period $\tau=1~\mathrm{ns}$. The Long-Kobayashi equations with noises are integrated with a fixed-step stochastic delay-differential-equation solver. A fourth-order Runge-Kutta scheme is used for the deterministic part and a Milstein update for the stochastic one. Each time step is $dt=0.5\tau_p = 2.9$~ps, and the total integration time is $500\tau = 500$ ns. The coupled lasers already reach a steady state in the final $50\tau$ from which we extract the lasing fields and frequencies. For a chosen set of coupling phases, the coupling budget increases gradually from $0$ to $60$~ns$^{-1}$ in 61 steps. In each step, the simulation runs for $500\tau$, and the final fields are used as initial values for the next step. The coupling budget is smoothly ramped over $150\tau$ to reduce transient effects. 
	
	For both uniform and proportional coupling, we scan the coupling phases $\phi_{nm}$ and find the coupling budget at the synchronization threshold. The instantaneous frequency of each laser is computed by the phase derivative of $E_m(t)$, and the steady-state frequency is extracted from temporal average over the final 50$\tau$ periods in 500$\tau$ simulation time. The frequency-locking threshold is given by the minimum coupling budget at which the frequency spread of all lasers is within $1\%$ of the initial spread. We compute the far-field emission contrast from all lasers to track phase locking, and find it occurs simultaneously with frequency locking.
	
	The locked laser phases are obtained from the complex field correlation function $g_{nm}^{(1)} = \langle E_m^*(t) \, E_n(t) \rangle$, where $\langle...\rangle$ represents average over time $t$. The argument of $g_{nm}^{(1)}$ gives the time-averaged relative phase between lasers $n$ and $m$ when they are synchronized. 
	
	\begin{figure*}[!ht]
		\centering
		\includegraphics[width=\linewidth]{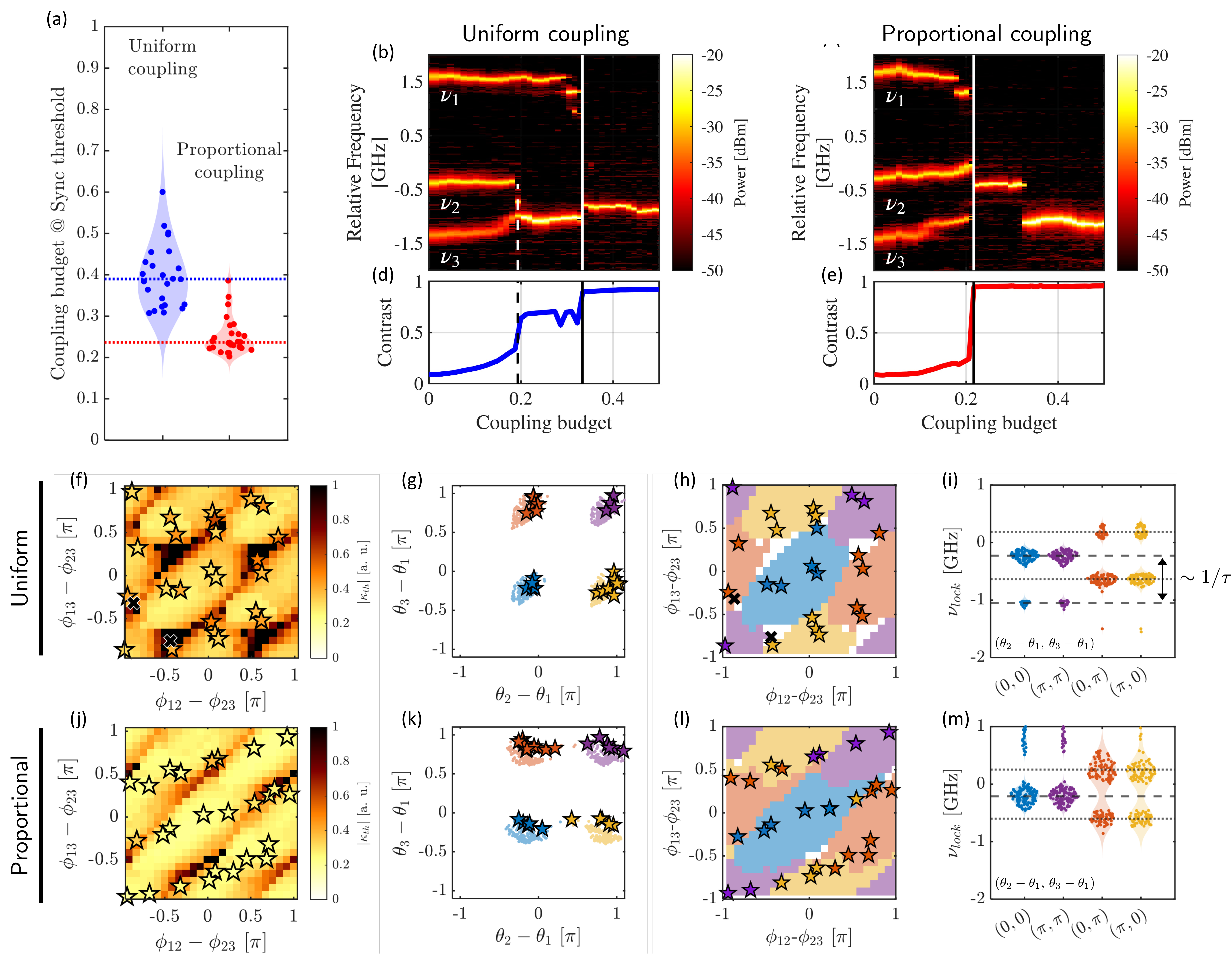}
		\caption{{\bf Uniform vs. proportional coupling at higher pump rate.} The pump current I = 1.6 mA is $r = 5$ times of the lasing threshold. (a) Experimentally measured synchronization thresholds for uniform (blue) and proportional (red) coupling of three VCSELs with different coupling phases (dots) in a swarmchart (dots), with colored background representing the probability density. Proportional coupling exhibits less threshold variation and lower medium threshold (dotted line) than uniform coupling.  (b, c) Representative experimental data of frequency locking for fixed coupling phases. Relative frequencies of three lasers with increasing coupling budget, showing sequential frequency locking (white dashed and solid lines) with uniform coupling (b) and simultaneous locking (white solid line) with proportional coupling (c). (d, e) Representative experimental data of phase locking for same coupling phases as in (b,c). Far-field fringe contrast showing two-step phase locking (black dashed and solid lines) with uniform coupling (d) and a single-step locking (black solid line) with proportional coupling (e). (f, j) Numerically calculated coupling budget at synchronization threshold for uniform and proportional coupling, respectively. Experimental data are overlaid as stars with inner color representing the measured coupling budget at synchronization threshold. Black crosses correspond to no synchronization observed experimentally up to the maximum coupling budget. (g, k) Relative phases of lasers 2 ($\theta_2$) and 3 ($\theta_3$) to laser 1 ($\theta_1$) at the synchronization threshold. Dots represent numerical results, and stars experimental data. The color encodes the locked-state phases: blue for $\theta_2-\theta_1 = \theta_3-\theta_1  = 0$, red for $\theta_2-\theta_1 = 0$ and $\theta_3-\theta_1 = \pi$, yellow for $\theta_2-\theta_1 = \pi$ and $\theta_3-\theta_1 = 0$, purple for $\theta_2-\theta_1 =  \theta_3-\theta_1 = \pi$. (h, l) Correspondence between relative phases of locked lasers and those of their coupling coefficients, from numerical simulation [colored regions with same color code as in (g, k)] and experiment (stars). The relative coupling phases are obtained experimentally. (i, m) Locked frequencies at the synchronization threshold for four sets of relative locked phases of three lasers [same color code as in (g, k)].}
		\label{fig:SI_pumpvariation}
	\end{figure*}
	
	\section{Additional experimental data}
	
	In the main text, we present experimental and numerical results of laser synchronization at one pumping level and one set of intrinsic frequencies. We have taken data at different pumping levels and with multiple frequency differences of lasers in the synchronization experiment to confirm the conclusions in the main text.
	
	\begin{figure*}[ht!]
		\centering
		\includegraphics[width=\linewidth]{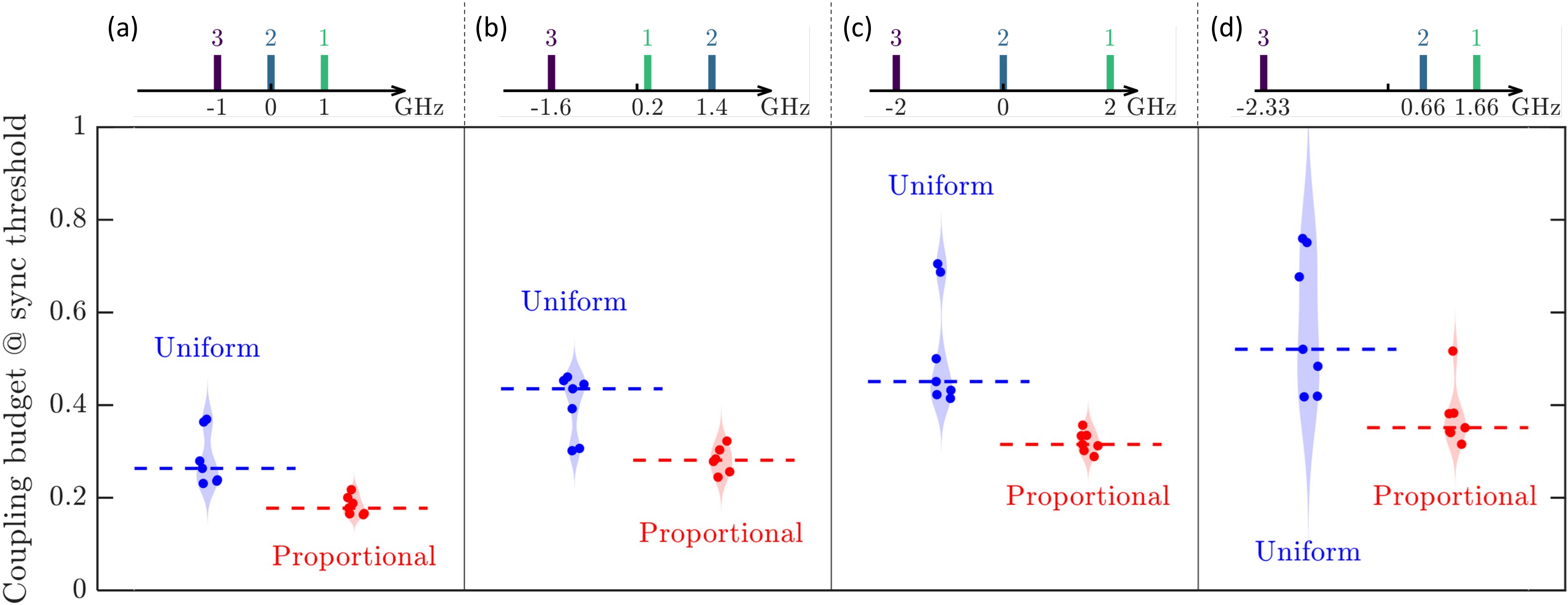}
		\caption{{\bf Uniform vs proportional coupling of three lasers with varying intrinsic frequencies.} Experimentally measured coupling budget at frequency/phase locking threshold for 4 sets of intrinsic frequency differences, as illustrated in the schematics for three VCSELs labeled 1, 2, 3. The maximum range of frequency difference ranges from 2 GHz to 4 GHz. For all sets of frequency differences, the uniform coupling (blue) exhibits a larger threshold variation with coupling phases and a higher median threshold value (dotted line) than proportional coupling (red). It is 1.48 times lower for panel (a), 1.54 times lower for panel (b), 1.43 times lower for panel (c) and 1.48 times lower for panel (d).}
		\label{fig:SI_freqvariation}
	\end{figure*}
	
	\subsection{Pump variation}
	
	Figure~\ref{fig:SI_pumpvariation} shows experimental and numerical results for three lasers with the same frequency detunings as in the main text, but at a pump current $I$ = 1.6 mA (5 times the lasing threshold, $r=5$). They are consistent with the results at $I$ = 1.0 mA (3 times the lasing threshold, $r=3$). For various coupling phases, the median coupling budget at the laser synchronization threshold for proportional coupling is 1.65 times lower than that of uniform coupling in  Fig.~\ref{fig:SI_pumpvariation}(a).  The lasing frequencies and far-field contrast in Figure~\ref{fig:SI_pumpvariation}(b) and (c) reveal sequential locking of three lasers with uniform coupling and simultaneous locking with proportional coupling.	We run simulations for $r=5$ and compare numerical results to experimental data in Figure~\ref{fig:SI_pumpvariation} (f-i) for uniform coupling and (j-m) for proportional coupling. The phase differences of locked lasers are close to $0$ or $\pi$, and the coupling phases dictate the locked laser frequency and relative phases, in agreement with the results in Figs. 3 and 4 of the main text.     
	
	\begin{figure*}[ht!]
		\centering
		\includegraphics[width=\linewidth]{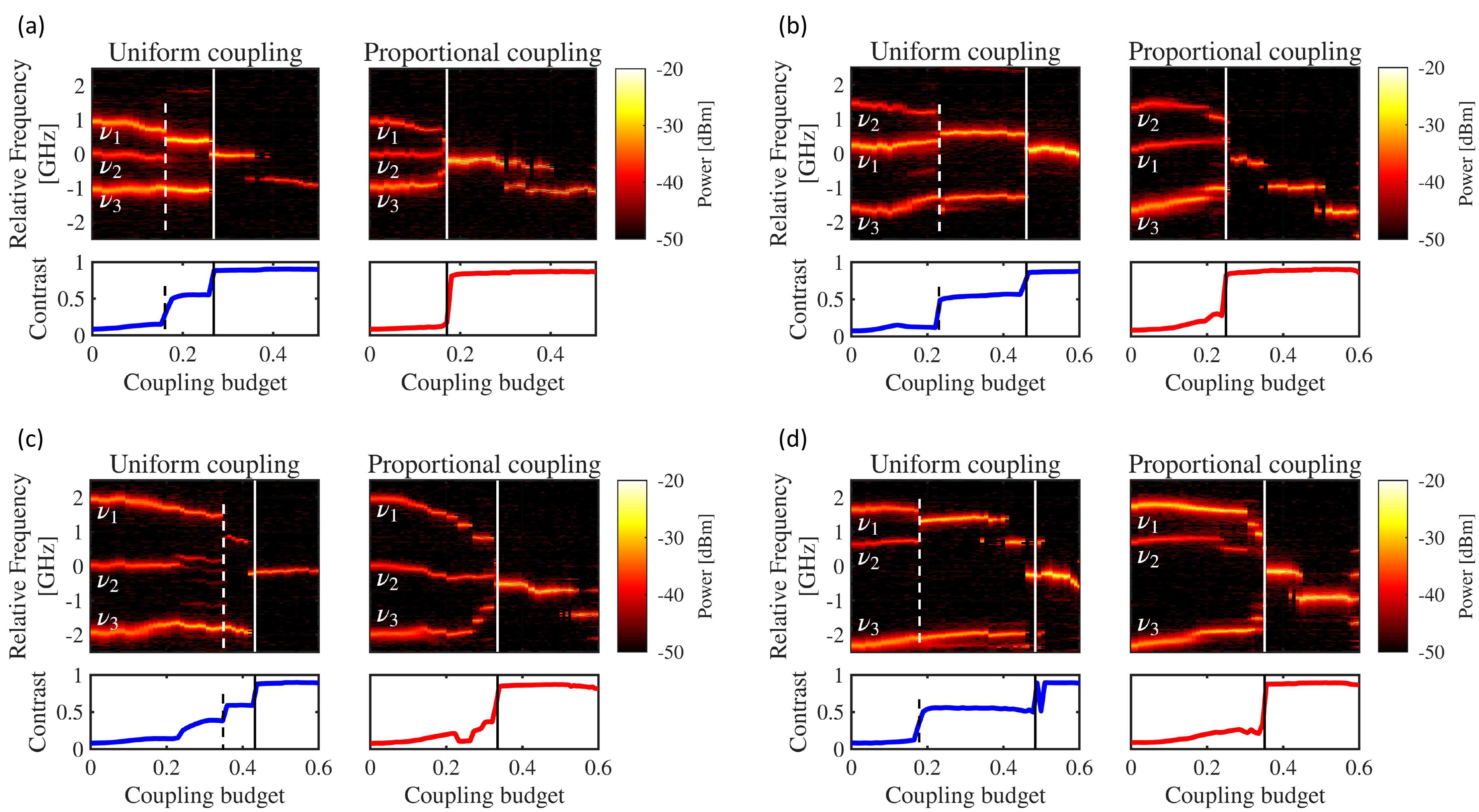}
		\caption{{\bf Representative examples of frequency and phase locking of three VCSELs with varying frequency differences.}  (a-d) corresponds to the four set of frequency differences in Fig.~\ref{fig:SI_freqvariation}:  $[1,\,0,\,-1]$ GHz in (a),  $ [0.2,\,1.4,\,-1.6]$ GHz in (b),  $[2,\,0,\,-2]$ GHz in (c), $ [1.66,\,0.66,\,-2.33]$ GHz in (d). Each panel compares uniform and proportional coupling for frequency locking in the heterodyne measurement of lasing frequencies and for phase locking in the measured far-field fringe contrast. The solid white line shows the frequency locking threshold. The dashed line indicates the two laser locking thresholds for uniform coupling. Despite different frequency detunings, the locking behaviors are similar.}
		\label{fig:SI_freqvariation_data}
	\end{figure*}
	
	\subsection{Intrinsic frequency differences}
	
	We experimentally vary the free-running frequencies of individual lasers by changing their pump currents. For each set of frequency differences, we compare proportional coupling with uniform coupling for varying coupling phases. From the pairwise frequency difference of uncoupled lasers, we scale their coupling amplitude for proportional coupling.
	
	Figure~\ref{fig:SI_freqvariation} shows a few examples. The average coupling budget required for synchronization increases with maximum frequency difference. The proportional coupling always has a lower median coupling budget than uniform coupling. On average, across all different configurations, the median coupling budget is 1.49 times lower for proportional than uniform coupling. 
	
	Representative scans of the frequencies and contrast as a function of the coupling budget are shown in Figure~\ref{fig:SI_freqvariation_data}. In all cases, the uniform coupling displays a sequential-locking behavior, while proportional coupling results in a collective transition to synchronization at lower coupling budget. Even when the intrinsic frequencies are evenly spaced as in Fig.~\ref{fig:SI_freqvariation_data}~(a, c), uniform coupling still exhibits sequential locking, because individual laser frequencies can be slightly changed by weak coupling before locking. 
	
	While changing the pump rate or the intrinsic frequency distribution modifies the locking threshold values and the coupling phases for most efficient locking, the main conclusions hold: proportional coupling lowers the mean frequency/phase locking threshold, and coupling phases determine how efficiently the available coupling budget is utilized for global synchronization.

	\clearpage

\end{document}